\documentclass[letterpaper, 10 pt, conference]{ieeeconf}
\usepackage{hyperref}
\usepackage{amsmath,amssymb,amsfonts}
\usepackage[noend]{algpseudocode}
\usepackage{algorithm}
\usepackage{array}
\usepackage[caption=false,subrefformat=parens]{subfig} 
\usepackage{textcomp}
\usepackage{stfloats}
\usepackage{verbatim}
\usepackage{graphicx}
\usepackage{cite}
\usepackage{booktabs}
\usepackage{multirow}
\newtheorem{remark}{Remark}
\usepackage[T1]{fontenc}
\usepackage{aecompl}
\IEEEoverridecommandlockouts                              

\begin{document}

\title{\LARGE \bf
Design and Experimental Test of \\ Datatic Approximate Optimal Filter in Nonlinear Dynamic Systems
}

\author{
    Weixian He$^{1}$,
    Zeyu He$^{1}$,
    Wenhan Cao$^{1}$,
    Haoyu Gao$^{1}$,
    Tong Liu$^{1}$,
    Bin Shuai$^{1}$,
    Chang Liu$^{3}$ and
    \\
    Shengbo Eben Li$^{2}$
    \thanks{
        $^{1}$Weixian He, Zeyu He, Wenhan Cao, Haoyu Gao, Tong Liu, Bin Shuai are with the School of Vehicle and Mobility, Tsinghua University, Beijing 100084, China. 
        (Emails: \texttt{\{hwx22,cwh19,ghy22,liu-t22\}@mails.tsinghua.edu.cn}, 
        \texttt{hezeyu1549@gmail.com},
        \texttt{shuaib@mail.tsinghua.edu.cn}, \texttt{lishbo@tsinghua.edu.cn}).
    }
    \thanks{
        $^{2}$Shengbo Eben Li is with the School of Vehicle and Mobility \& College of AI, Tsinghua University, Beijing 100084, China.
        (Email: \texttt{lishbo@tsinghua.edu.cn}).
    }
    \thanks{
        $^{3}$Chang Liu is with the College of Engineering, Peking University, Beijing 100871, China. 
        (Email: \texttt{changliucoe@pku.edu.cn}).
    }
    \thanks{ 
        This work was supported by NSF China under 52221005. It is also partially supported by Tsinghua University-Toyota Joint Research Center for AI Technology of Automated Vehicle, and Tsinghua University Initiative Scientific Research Program.\textit{Corresponding author: Shengbo Eben Li (lishbo@tsinghua.edu.cn)}
    }
}
\maketitle

\begin{abstract}
Filtering is crucial in engineering fields, providing vital state estimation for control systems. However, the nonlinear nature of complex systems and the presence of non-Gaussian noises pose significant challenges to the performance of conventional filtering methods in terms of estimation accuracy and computational efficiency. In this work, we present a data-driven closed-loop filter, termed datatic approximate optimal filter (DAOF), specifically designed for nonlinear systems under non-Gaussian conditions. We first formulate a Markovian filtering problem (MFP), which inherently shares a connection with reinforcement learning (RL) as it aims to compute the optimal state estimate by minimizing the accumulated error. To solve MFP, we propose DAOF, which primarily incorporates a trained RL policy and features two distinct structural designs: DAOF-v1 and DAOF-v2. Designed for systems with explicit models, DAOF-v1 combines prediction and update phases, with the RL policy generating the update value. Meanwhile, DAOF-v2 bypasses system modeling by directly outputting the state estimate. Then, we utilize an actor-critic algorithm to learn the parameterized policy for DAOF. Experimental results on a 2-degree-of-freedom (2-DOF) vehicle system, equipped with explicit system models, demonstrate the superior accuracy and computational efficiency of DAOF-v1 compared to existing nonlinear filters. Moreover, DAOF-v2 showcases its unique ability to perform filtering without requiring explicit system modeling, as validated by a 14-DOF vehicle system.

\end{abstract}


\section{Introduction}
Filtering plays a pivotal role in engineering domains by offering essential state estimation for control systems.  Bayesian filtering, underpinning various filtering techniques, utilizes Bayes’ theorem to refine the posterior distribution of the system state. This process allows for the estimation of the system state through the application of a chosen criterion \cite{sarkka2023bayesian}. The seminal Kalman filter (KF) represents a specific instance of Bayesian filtering, known for its optimality in linear systems corrupted by Gaussian noise, where it minimizes the mean square error \cite{kalman1960new}. For nonlinear dynamic systems, extended Kalman filter (EKF) has gained widespread use. EKF approximates nonlinear stochastic models by linearizing them through a first-order Taylor series expansion \cite{smith1962application}. However, this linearization inherent to EKF results in suboptimal performance due to the neglect of higher-order terms, potentially causing algorithmic instability. An alternative approach is unscented Kalman filter (UKF), which uses a set of sigma points to capture the mean and covariance of the posterior distribution without direct linearization of the model \cite{julier1997new}. Nevertheless, UKF may experience divergence when the system model significantly deviates from actual dynamics or when the noise characteristics are asymmetric \cite{perea2007nonlinearity}. In scenarios with strong nonlinearity or non-Gaussian noises, the effectiveness of both EKF and UKF tends to degrade substantially. 

Particle filter (PF),  rooted in Monte Carlo sampling principles \cite{liu1998sequential}, offers a more precise and theoretically sound approach to a wide range of filtering problems than the variants of KF. However, PF requires the propagation and resampling of a large number of particles at each iteration, which leads to a significant increase in computational overhead. This computational burden poses a challenge for its deployment in real-time estimation applications. In contrast to PF, which approximates the full posterior distribution with particles, moving horizon estimation (MHE) focuses on obtaining the maximum a posteriori (MAP) estimate by iteratively minimizing an estimation cost function across a moving time horizon \cite{rao2001constrained}. Similar to PF, the improved accuracy of MHE comes at the cost of increased computational effort, especially in situations with complex, high-dimensional state spaces.

The swift progress in deep learning techniques in recent times has spurred the growth of learning-based filtering approaches \cite{al2019deep, jin2021new, bai2023state}. For example, researchers have integrated deep learning into estimation by merging structural state-space models with recurrent neural networks within the KF framework, yielding promising results in numerical simulations \cite{revach2022kalmannet, chen2021dynanet}. Nevertheless, these methods are often criticized for their lack of robust theoretical underpinnings, which complicates the establishment of their equivalence or near-equivalence to optimal filtering algorithms. The variational Bayes filter, another learning-based method, employs deep neural networks (DNNs) in a recursive framework to model the variational distribution of process noises, contingent on new observations and prior estimates \cite{karl2017deep}. This approach benefits from a solid theoretical basis through variational inference and has shown reliable estimation performance in real-world experiments. Despite this, the assumption of Gaussian noise and the need for precise modeling limit its application in diverse scenarios where noise types are unknown or system models are imprecise. Efforts to approximate MHE using DNNs have also been documented \cite{allan2019moving}. However, these methods often fail in accuracy because they focus on the mode of the state posterior rather than the full distribution, and they also face difficulties in determining the arrival cost in MHE during training.

A shared challenge among the previously mentioned filtering techniques is their heavy reliance on the accuracy of system models and parameters \cite{kalman1960new, smith1962application, julier1997new, perea2007nonlinearity, liu1998sequential, rao2001constrained}, or at least a structural representation of the system \cite{revach2022kalmannet, chen2021dynanet}, a characteristic commonly associated with modelic (i.e., model-driven) filtering. This dependency on explicit and accurate models severely restricts their utility in complex systems with inherently difficult dynamics, such as complex vehicle systems with intricate dynamics \cite{li2023maximizing}. Consequently, there is a pressing demand for a shift from modelic filtering to datatic (i.e., data-driven) filtering, where the dynamics of systems are inferred from data samples rather than relying on explicit system models, thereby diminishing the model dependency.


Model-free reinforcement learning (RL) has gained prominence as a paradigmatic strategy for datatic control challenges \cite{shakya2023reinforcement, aliramezani2022modeling}. This approach offers a viable alternative in complex situations where deriving a state-space model explicitly is difficult. Furthermore, RL excels in tackling high-dimensional issues using high-capacity approximation functions, such as DNNs, for representing policy and value functions \cite{arulkumaran2017deep, wang2022deep}. The application of RL to the field of filtering showcased remarkable capabilities \cite{RN10}, yet its cost function was initially designed for zero-mean Gaussian noise. A recent innovation in RL-based filtering has shown enhanced accuracy and computational efficiency over conventional filters such as UKF and PF in nonlinear systems with non-Gaussian noises \cite{cao2021reinforced}. Nonetheless, this technique still requires explicit system modeling, rendering it modelic.


To overcome the aforementioned challenges, this paper proposes a datatic approximate optimal filter (DAOF) for state estimation, which is represented by DNNs and trained through RL techniques. The main contributions of this paper are delineated as follows:
\begin{itemize}
    \item We formulate a Markovian filtering problem (MFP) by utilizing a discounted accumulated error metric as the evaluation criterion and present its standard form. To solve MFP, we derive DAOF adapted for nonlinear systems affected by non-Gaussian noises.  
    \item  We develop two distinct policy structures for DAOF, namely DAOF-v1 and DAOF-v2, for the cases with and without explicit system modeling to address MFP. We employ an RL algorithm in actor-critic framework to train the policy for DAOF, which is parameterized by a multi-layer neural network.
    \item We conduct experiments on two vehicle systems to validate the efficacy of DAOF. DAOF-v1 showcases superior accuracy and computational efficiency compared to traditional nonlinear filters. The unique capability of DAOF-v2 to perform filtering without requiring explicit system models is also demonstrated.
     
\end{itemize}

\section{Preliminaries}
Filtering is the technique by which hidden state variables are inferred from noisy measurement data \cite{anderson2012optimal}. Essentially, a filter serves as a mapping of observed information, where the input consists of a sequence of noisy measurements from time $1$ to time $t$, denoted as $y_{1:t}$, and the output provides the estimated state, represented by $\hat{x}_t$. 
We consider a discrete-time stochastic system, characterized by two state space equations: (1) transition equation, which depicts the system behaviors with process uncertainties, and (2) measurement equation, which describes the characteristics of imperfect sensors:
\begin{equation}
\label{eq.system}
\begin{aligned}
x_{t+1}&=f(x_{t})+\xi_{t}, \\
y_{t}&=g(x_{t})+\zeta_{t},\\
x_{0}&\sim p(x_{0}),\\
t &\geq 0,
\end{aligned}
\end{equation}
where $x_{t}\in\mathbb{R}^{n}$ is the state, $y_{t}\in\mathbb{R}^{m}$ is the measurement, $f(\cdot)$ and $g(\cdot)$ are the time-invariant transition and measurement functions,  respectively, and $\xi_{t}\in\mathbb{R}^{n}$ is the process noise and $\zeta_{t}\in\mathbb{R}^{m}$ is the measurement noise. The stochastic system is assumed to satisfy the following conditions \cite{anderson2012optimal}:
\begin{enumerate}
\item The process noise $\xi_{0:\infty}$ is independent and identically distributed (i.i.d), which obeys a distribution $p_{\xi}(\xi_{t})$.
\item The observation noise $\zeta_{0:\infty}$ is i.i.d, which obeys a distribution $p_{\zeta}(\zeta_{t})$.
\item The noises $\xi_{0:\infty}$ and $\zeta_{0:\infty}$ are independent of each other.
\item The distribution of $x_{0}$ is independent of  $\xi_{0:\infty}$ and $\zeta_{0:\infty}$.
\end{enumerate}

\begin{figure}[!h]
    \centering
    \includegraphics[width=0.5\textwidth]{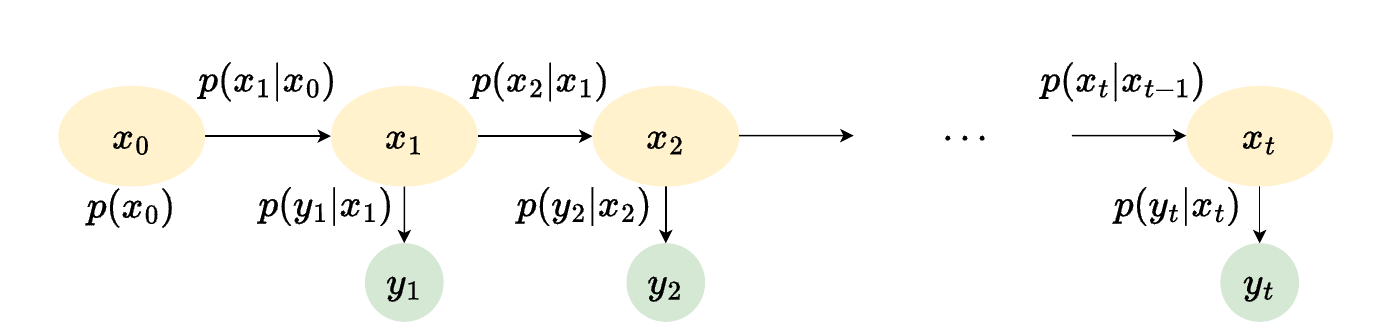}
    \caption{Hidden markov model for filtering problems.} 
    \label{fig:hmm}
\end{figure}

In addition to aforementioned descriptions, as shown in Fig. \ref{fig:hmm}, we can also utilize the hidden Markov model to characterize stochastic systems, in which the measurement $y_{t}$ is known and the true state $x_{t}$ is the latent variable to be inferred at every time step.
The transition equation is equivalently replaced by the one-step transition probability $p(x_{t} | x_{t-1})$, while the measurement equation is represented by the one-step output probability $p(y_{t} | x_{t})$.

For the filtering problem, we consider two distinct scenarios. In the first situation, the transition function $f(\cdot)$ or the one-step transition probability $p(x_{t} | x_{t-1})$, the observation function $g(\cdot)$ or the one-step output probability $p(y_{t} | x_{t})$, the initial state distribution $p(x_0)$, the noise distributions $p_{\xi}(\xi_{t})$ and $p_{\zeta}(\zeta_{t})$, as well as the measurements $y_t$ are all available. In contrast, the second scenario presents a more limited set of information, encompassing the initial state distribution $p(x_0)$, the measurements $y_t$ and the true state values $x_t$ (acquired from a simulator or high-precision sensors). Standard filters such as KF, EKF, UKF are all designed for the first case. However, filters designed for the second, more constrained scenario are less common. We are focused on creating filters that are suitable for both of these mentioned scenarios by utilizing RL algorithms. To this end, we have constructed MFP by using discounted accumulated error as the criterion:
\begin{equation}
\label{eq:optimal problem}
\hat{x}^*_{t:\infty}=\mathop{\arg\min}_{\hat{x}_{t:\infty}}\mathop{\mathbb{E}}\left\{\sum_{k=t}^{\infty} \gamma^{k-t} \phi(x_{k}, \hat{x}_{k})\mid h_{t}
\right\},
\end{equation}
where $h_{t}
=(\hat{x}_{0}, y_{1}, \hat{x}_{1}, y_{2}, \ldots, \hat{x}_{t-1}, y_{t})
$ is the historical information containing all past estimations and observations, $\phi(\cdot, \cdot)$ is the risk function, which is to measure the difference between the true state and the estimate, and $\gamma\in[0,1)$ is the discount factor. The expectation is taken over $x_{t:\infty}$. 

\begin{remark}
    In the second scenario, we note that the true state values need to be known, meaning that during the training process, the true states are required to calculate the loss. However, once training is complete, the filter can be applied online without needing access to the true states.
\end{remark}

\section{Method}
In this section, we initially formulate the standard form of MFP and outline its connections with the fundamentals of RL. Subsequently, we derive DAOF, which is particularly suitable for nonlinear systems influenced by non-Gaussian noises to solve MFP. To tackle the challenges associated with two distinct filtering scenarios, distinguished by the availability or lack of an explicit system model, we craft two tailored policy architectures. Then, we employ an RL algorithm in actor-critic framework to train the policy for DAOF, which is parameterized by a multi-layer neural network.
\subsection{Standard form of MFP}
To apply RL algorithms, it is crucial  to ensure that the system adheres to the Markov property. Given the established assumptions about the stochastic system, the subsequent state $x_{t+1}$ is solely dependent on the 2-tuple $(x_t, \xi_t)$. This dependency indicates that the environmental dynamics indeed satisfies the Markov property. However, the current state $x_t$ cannot be directly ascertained from observations alone due the inherent observation uncertainty. A practical strategy is to treat the historical information $h_t$ as an augmented "state". By adopting this approach, it becomes  possible to iteratively update this expanded “state” as follows:
\begin{equation}\nonumber
    h_{t+1}=\text{Update}\left\{h_t, \hat{x}_{t}, y_{t+1}\right\}.
\end{equation}
Apparently, the set $h_t$ constitutes the union encompassing all elements from $h_1$ to $h_t$: 
\begin{equation}\nonumber
    h_t = \bigcup\limits_{i=1}^{t} h_i.
\end{equation}
From this perspective, the ``state" that owns all the historical information holds the Markov property:
\begin{equation}\nonumber
    p(h_{t+1}\mid h_{t})=p(h_{t+1}\mid h_{1}, h_{2}, \ldots, h_{t} ).
\end{equation}
Subsequently, we can align in the components of the filtering problem with the elements of RL framework. Consider the standard RL setup for MFP, the state is denoted  as $s_t=h_t$. Action is the estimate following a filtering policy, which can be described as $a_t=\hat{x}_{t}=\pi({h_t})$. The cost function is formulated as $l_t=l(x_{t},\hat{x}_{t})=\|x_{t}-\hat{x}_{t}\|_{2}^{2}$, where the true state $x_t$ is expressible as a function of $h_t$ due to the system properties defined in \eqref{eq.system}. 
Moreover, the transition probability $p(h_{t+1}|h_t,\hat{x}_t)$ maps a given state-action tuple $(s_t,a_t)$ to the probability distribution over $s_{t+1}$. 

Building upon this foundation, we can construct the action-value function as follows:
\begin{equation}
\label{eq:action-value function}
    Q(h_t, \hat{x}_t) = \mathop{\mathbb{E}}\left\{\sum_{k=t}^{\infty} \gamma^{k-t}\|e_{k}\|_{2}^{2} \mid h_t, \hat{x}_t\right\}.
\end{equation}
This problem formulation aims to find the optimal estimate with the minimized value function, i.e.,
\begin{equation}
\label{eq.objective-model}
\begin{aligned}
\hat{x}^*_{t:\infty}&=\mathop{\arg\min}_{\hat{x}_{t:\infty}}Q(h_{t},\hat{x}_{t}).
\end{aligned}
\end{equation}
We refer to the problem \eqref{eq.objective-model} for systems in \eqref{eq.system} as the standard form of MFP.

The typical approach to obtaining the optimal solution for an RL problem is to derive its Bellman equation \cite{bellman1966dynamic}. Consequently, we aim to derive the Bellman equation for MFP in order to solve the optimal filtering policy. Recall that the state $h_t$ has the Markov property, and the cost function $l_t$ is structurally separable over time steps. These two properties can be utilized to separate the current-time utility function from its successors \cite{li2023reinforcement}. Then, we can derive the Bellman equation by Bellman's principle of optimality:
\begin{equation}
    \begin{aligned}
        Q^{*}(h_{t}, \hat{x}_{t})=&\mathop{\mathbb{E}}_{{h_{t+1}\sim p(\cdot\mid h_t,\hat{x}_{t})}}\left\{\|e_{t}\|_{2}^{2}\right\}
        \\&+\gamma \min_{\hat{x}_{t+1} 
        }\mathop{\mathbb{E}}_{{h_{t+1}\sim p(\cdot\mid h_t,\hat{x}_{t})}}\left\{Q^{*}(h_{t+1}, \hat{x}_{t+1}) \right\}.
    \end{aligned}
\end{equation}

According to \eqref{eq.objective-model}, solving MFP requires obtaining the optimal estimates at all future time steps, which is clearly impractical. However, by leveraging the Bellman equation derived above, we only need to seek the optimal estimate at the current time point at each step, thereby rendering the problem more tractable.

\begin{remark}
Employing a proof methodology analogous to that presented in \cite{cao2021reinforced}, we demonstrate that the optimal solution of MFP in linear Gaussian systems coincides with the stationary Kalman filter, thus validating the correctness of this problem formulation.
\end{remark}

\subsection{Policy Structure of DAOF}
Drawing inspiration from the predictor-updater structure of KF, for the first kind of filtering scenario where the dynamic system is explicitly modeled, we have structured DAOF-v1 as outlined below:
\begin{equation}\label{equation:DAOF x_hat}
\begin{aligned}   \hat{x}_{t}=f({\hat{x}_{t-1}})+\pi(h_{t}),
\end{aligned}
\end{equation}
where $f(\cdot)$ is the prediction component that is known from the system model and $\pi(
h_{t})$ is the update component to be learned.  Another form of DAOF-v1 is as follows:
\begin{equation}
\label{eq:g}
    \hat{x}_t = f\left(g^{-1}(y_{t-1})\right) + \pi(h_{t}).
\end{equation}
The structure in \eqref{equation:DAOF x_hat} only utilizes the information from the transition model, while the structure in \eqref{eq:g} fully utilizes both transition and observation models. Nevertheless, the latter method’s reliance on inverse operations limits its applicability in intricate nonlinear systems. Therefore, this paper focuses primarily on the former structure.

For the second filtering challenge, where the explicit formulation of system models is unattainable, a model-free policy is crucial. Taking vehicle dynamics as an example, achieving accurate state estimation necessitates the development of sophisticated models, such as the 14-DOF vehicle model\cite{li2023maximizing}. Traditional solving techniques like Euler discretization may fail in providing the necessary precision, and the preference for higher-order methods, including the fourth-order Runge-Kutta technique, implies that a direct representation of transition and observation equations is often impractical.
In light of these challenges, we design DAOF-v2 to directly estimate the state as follows:
\begin{equation}
    \hat{x}_{t} = \pi(
    h_{t}).
\end{equation}
DAOF with both structures can be characterized as a closed-loop filter. Here, the state estimates from prior time steps are integrated into the policy input, which in turn provides a more comprehensive grasp of the state being estimated, thereby augmenting the overall filtering performance.

After selecting the policy structure for DAOF, we employ the widely utilized policy iteration techniques in RL to find the nearly optimal value and policy functions. Policy iteration involves two iteration procedures: 1) policy evaluation (PEV) and 2) policy improvement (PIM).

Given a filtering policy $\pi$, PEV seeks to numerically solve its corresponding action-value function $Q^{\pi}(h_{t},\hat{x}_{t})$: 
\begin{equation}
    \begin{aligned}
        Q^{\pi}_{j+1}(h_{t}, \hat{x}_{t})&=\mathop{\mathbb{E}}_{{h_{t+1}\sim p(h_{t+1}\mid h_t,\pi(h_{t}))}}\left\{\|e_{t}\|_{2}^{2}\right\}
        \\&+\gamma \mathop{\mathbb{E}}_{\substack{
        h_{t+1}\sim p(h_{t+1}\mid h_t,\pi(h_{t}))}}\left\{Q^{\pi}_{j}(h_{t+1}, \pi(h_{t+1})) \right\}.
    \end{aligned}
\end{equation}

PIM is derived as:
\begin{equation}
\label{PIM-free}
\begin{aligned}
\pi^{k+1}(
h_{t})=\mathop{\arg\min}_{\pi} 
Q_{\infty}^{\pi^k}(h_{t}, \hat{x}_{t}),
\end{aligned}
\end{equation}
where $Q^{\pi^k}_{\infty}$ is the stationary action-value function of the estimator $\pi^{k}$, which is obtained through iterations in PEV.

\begin{remark}
    In the formulation of policy structures and the derivation of policy iteration process, we eschew any assumptions about the nature of system noises, including the common assumption of Gaussian distribution. This flexibility allows DAOF to potentially manage non-Gaussian noises.
\end{remark}



\subsection{Training Algorithm Design}

By far, we have formulated MFP and designed the filter using complete historical information, thereby ensuring Markov property. However, the pursuit of perfect numerical computation becomes increasingly difficult as the dimensionality of historical information expands over time, whereas RL algorithms are constrained to accommodate only a fixed-length state vector. Consequently, we employ a fixed-length sliding window, denoted as $h^{N}_t=(\hat{x}_{t-1}, y_t, \hat{x}_{t-2}, y_{t-1}, \ldots, \hat{x}_{t-N}, y_{t-N+1})$ to approximately represent historical information. The selection of the length of the sliding window allows for a flexible trade-off between real-time processing constraints and accuracy requirements. Subsequently, we can utilize an RL algorithm in actor-critic framework to learn the policy for DAOF. 

Multi-layer neural networks are used to represent the value function and the filtering policy to obtain DAOF. The value network, also called critic, is denoted as:
\begin{equation}
\begin{aligned}
    Q(h^{N}_t, \hat{x}_t) &\cong Q(h^{N}_t, \hat{x}_t;w),
\end{aligned}
\end{equation}
where $w$ is the parameter.

The critic loss is derived as
\begin{equation}
    \begin{aligned}
        J_{\rm{critic}}=\mathop{\mathbb{E}}_{
        h^{N}_t, \hat{x}_t, h^{N}_{t+1} \sim \mathcal{D}_{\rm{Replay}}}\Big\{\frac{1}{2}\big[ l_{t}&+\gamma Q(h^{N}_{t+1}, \hat{x}_{t+1} ; w) \\
        &-Q(h^{N}_t, \hat{x}_t ; w)\big] ^{2}\Big\},
    \end{aligned}
\end{equation}
where $\mathcal{D}_{\rm{Replay}}$ is the replay buffer, and $l_t=l(h^{N}_{t},\pi(h_{t}))=\mathop{\mathbb{E}}
\left\{\|e_{t}\|_{2}^{2}\right\}$ is the cost function.

The parameter $w$ can be optimized with the following gradient:
\begin{equation}\label{equation:gradient of critic model}
\begin{aligned}
\frac{\partial J_{\rm {critic }}}{\partial w}=&-\mathop{\mathbb{E}}_{h^{N}_t, \hat{x}_t, h^{N}_{t+1} \sim \mathcal{D}_{\rm{Replay}}}\Big\{\big[l_{t}+\gamma Q(h^{N}_{t+1}, \hat{x}_{t+1} ; w) \\
&-Q(h^{N}_t, \hat{x}_t ; w)\big]\frac{\partial Q(h^{N}_t, \hat{x}_t ; w)}{\partial w}\Big\},
\end{aligned}
\end{equation}

The filter network is also called the actor, denoted as:
\begin{equation}
\pi(
h^{N}_t) \cong \pi(
h^{N}_t ; \varphi),
\end{equation}
where $\varphi$ is the parameter. According to \eqref{PIM-free}, a better policy network can be obtained by minimizing the following actor loss:
\begin{equation}
\begin{aligned}
    J_{\rm{actor}}&=\mathop{\mathbb{E}}_{h^{N}_t, \hat{x}_t \sim \mathcal{D}_{\rm{Replay}}}\left\{Q(h^{N}_t, \hat{x}_t ; w)\right\}.
\end{aligned}
\end{equation} 

The update gradient of actor can be expressed as:
\begin{equation}
\label{equation:gradient of actor free}
\begin{aligned}
    \frac{\partial J_{\rm{actor}}}{\partial \varphi}=\mathop{\mathbb{E}}_{h^{N}_t, \hat{x}_t \sim \mathcal{D}_{\rm{Replay}}}\Big\{
    \frac{\partial Q(h^{N}_{t}, \hat{x}_{t};w)}{\partial \hat{x}_{t}} 
    \frac{\partial \pi}{\partial \varphi}
    \Big\},
\end{aligned}
\end{equation}

\begin{algorithm}[H]
\caption{Training algorithm for DAOF}\label{alg:DAOF}
\begin{algorithmic}
\State Initialize parameters      $\varphi_0$, $w_0$, $\alpha$, $\beta$  
\State Initialize state $h_0\in\mathcal{S}$  
\Repeat
\State \quad Rollout with estimator $\pi$ from $h^{N}_{t}$
\State \quad Receive and store $h^{N}_{t+1}$
\State \textbf{PEV step}:
\State Calculate critic gradient \eqref{equation:gradient of critic model} 
\State Update value function with:
$
w_{k+1}=w_k-\alpha\frac{{\partial J}_{\rm{critic}}}{\partial {w}}
$
\State \textbf{PIM step}:
\State Calculate actor gradient \eqref{equation:gradient of actor free} 
\State Update estimator with:
$
\varphi_{k+1}=\varphi_k-\beta\frac{{\partial J}_{\rm{actor}}}{\partial{\varphi}}
$
\Until Convergence
\end{algorithmic}
\end{algorithm}

\begin{remark}
In this section, we provide an actor-critic algorithm framework, which means that any RL algorithm with a similar structure can be applied to this framework. In the experiments presented in this paper, we utilize DSAC-T, which has achieved leading performance in mainstream tests \cite{duan2023dsact}.
\end{remark}

\begin{figure*}[htbp]
 \centering
 \includegraphics[width=0.8\linewidth]{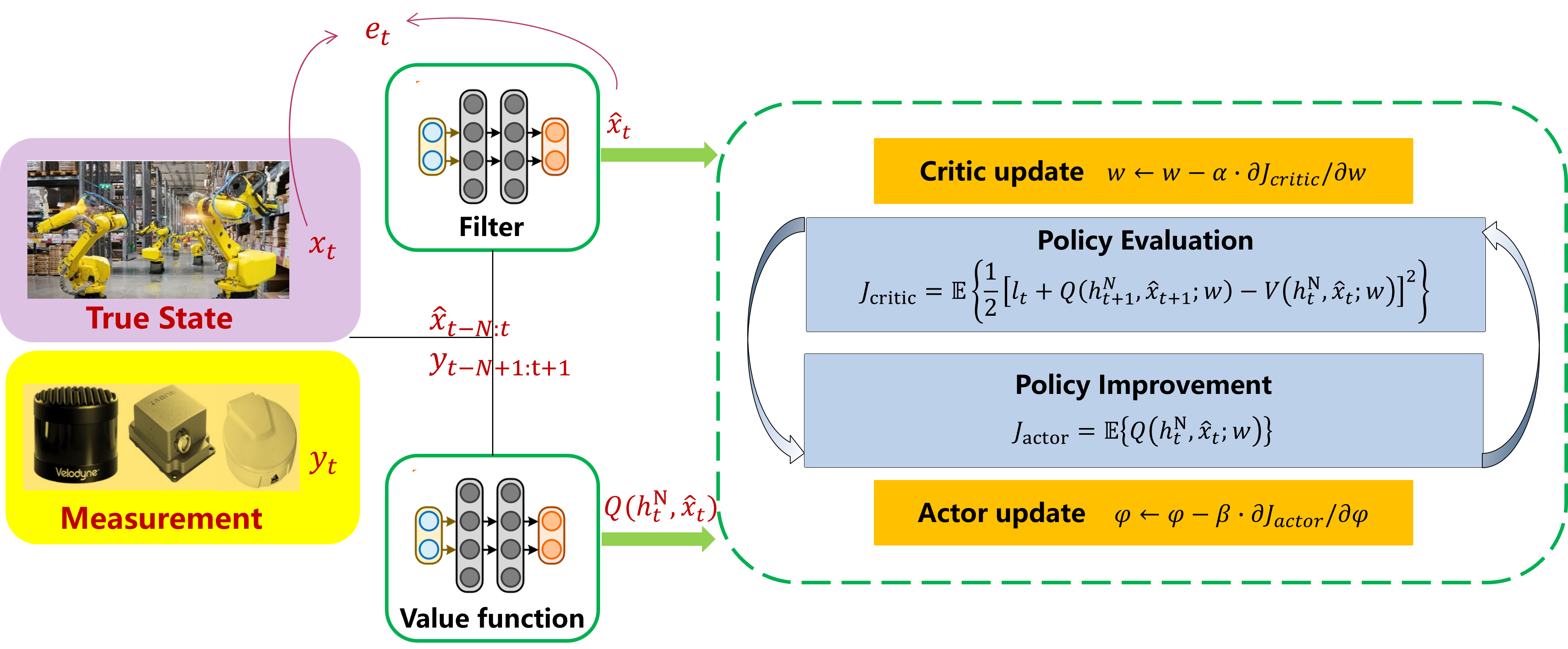}
 \caption{Training procedure of DAOF. The learned filter can be applied online.}
 \label{fig:algorithm}
\end{figure*}

\section{Experiments}
\label{sec:experiment}
In this section, we evaluate the performance of our proposed methods on two nonlinear vehicle systems with non-Gaussian noises. Notably, we incorporated Gaussian mixture noise for the process noise to capture the potential multimodal nature of the system dynamics. The Gaussian mixture model is theoretically capable of approximating any continuous probability distribution, which is particularly advantageous for mimicking real-world scenarios \cite{pfeifer2019expectation}. For observation noises, we employ a Laplace distribution, known for its heavy tails, to effectively represent outliers in observations that may arise from sensor faults or extrinsic disturbances \cite{neri2021approximate}. We carry out a comparative study involving our proposed DAOF-v1 and DAOF-v2, alongside several conventional methods, including UKF, PF, and a supervised learning-based filter (SLF) constructed with a multi-layer network that has a structure similar to the policy network of DAOF. The performance of these filters is examined through Monte Carlo simulations, each consisting of 500 time steps and repeated 100 times with various initial conditions. The evaluation metrics include the Root Mean Square Error (RMSE) between the estimated and actual states, as well as the computational cost, which is defined as the average time consumed per estimation iteration. The experimental analysis was conducted on a computing system equipped with a 3.4 GHz AMD Ryzen 7 5800 8-Core Processor. 

\subsection{Experiment I: Estimation of a 2-DOF Vehicle Model}
Consider the following vehicle dynamics \cite{bakker1987tyre}:
\begin{equation*}
    \begin{bmatrix}
        \delta_{t+1}\\
        \Omega_{t+1}
    \end{bmatrix}=\begin{bmatrix}
        \delta_{t}\\
        \Omega_{t}
    \end{bmatrix} + \begin{bmatrix}
        f_1(\delta_t, \Omega_t, u_t)\\
        f_2(\delta_t, \Omega_t, u_t)
    \end{bmatrix}\Delta \tau + \xi_t,
\end{equation*}
where $\delta_t$ and $\Omega_t$ are the side slip angle and the yaw rate of the vehicle, respectively. The steering angle $u_t$ serves as the control input, causing the vehicle to travel under a sinusoidal driving condition. Considering tire characteristics, the model shows significant nonlinearity. The measurement $y_t$ is given by:
\begin{gather*}
    y_t = [a_{y,t}, \Omega_t]^\top + \zeta_t,\\
    a_{y,t}= u_tf_1(\delta_t, \Omega_t) + u_t\Omega_t,
\end{gather*}
where $a_{y,t}$ is the lateral acceleration. 
As mentioned before, the process noise $\xi_t$ follows a Gaussian mixture model, while the measurement noise $\zeta_t$ follows a Laplace distribution. 

\begin{figure}[!h]
    \centering
    \subfloat{
        \includegraphics[width=0.225\textwidth]{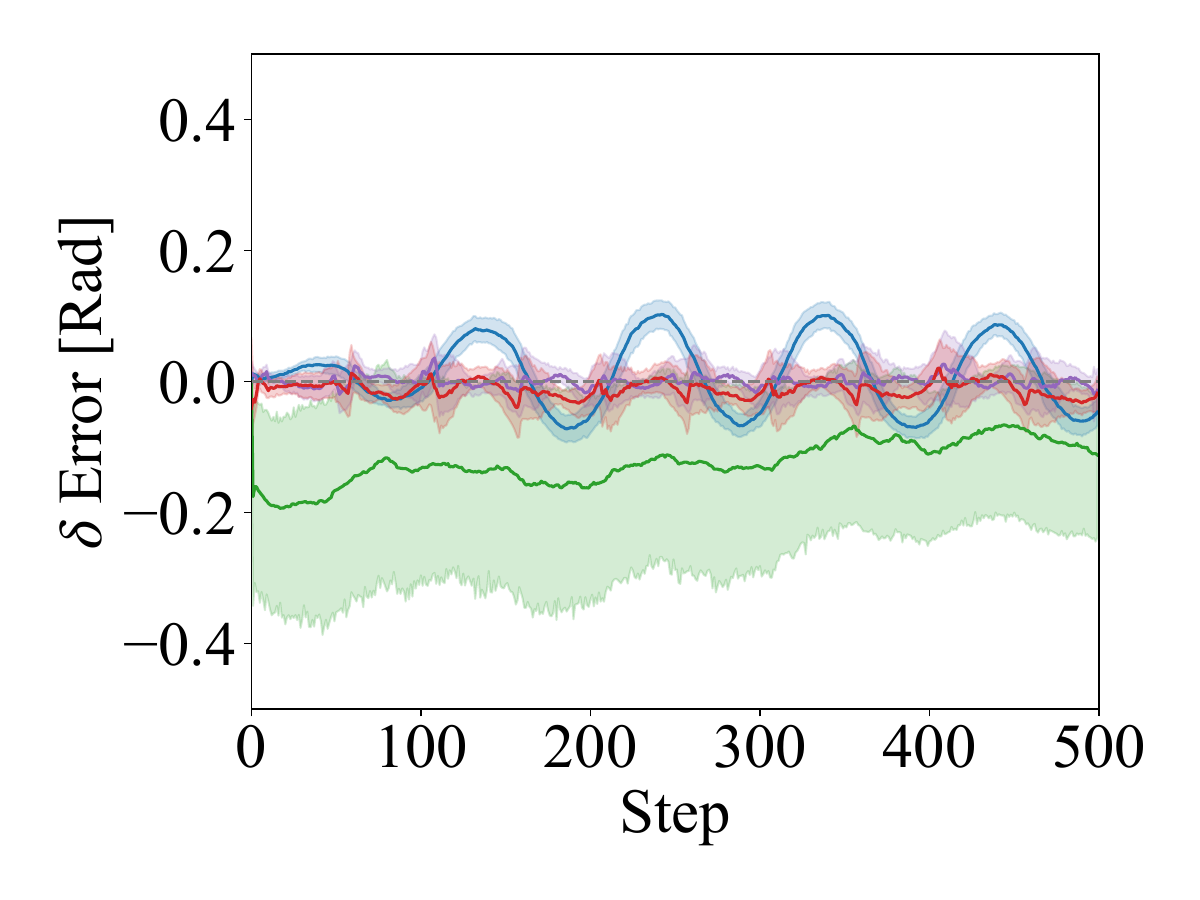}
        \label{sensor_net:1}
    }
    \hfill
    \subfloat{
        \includegraphics[width=0.225\textwidth]{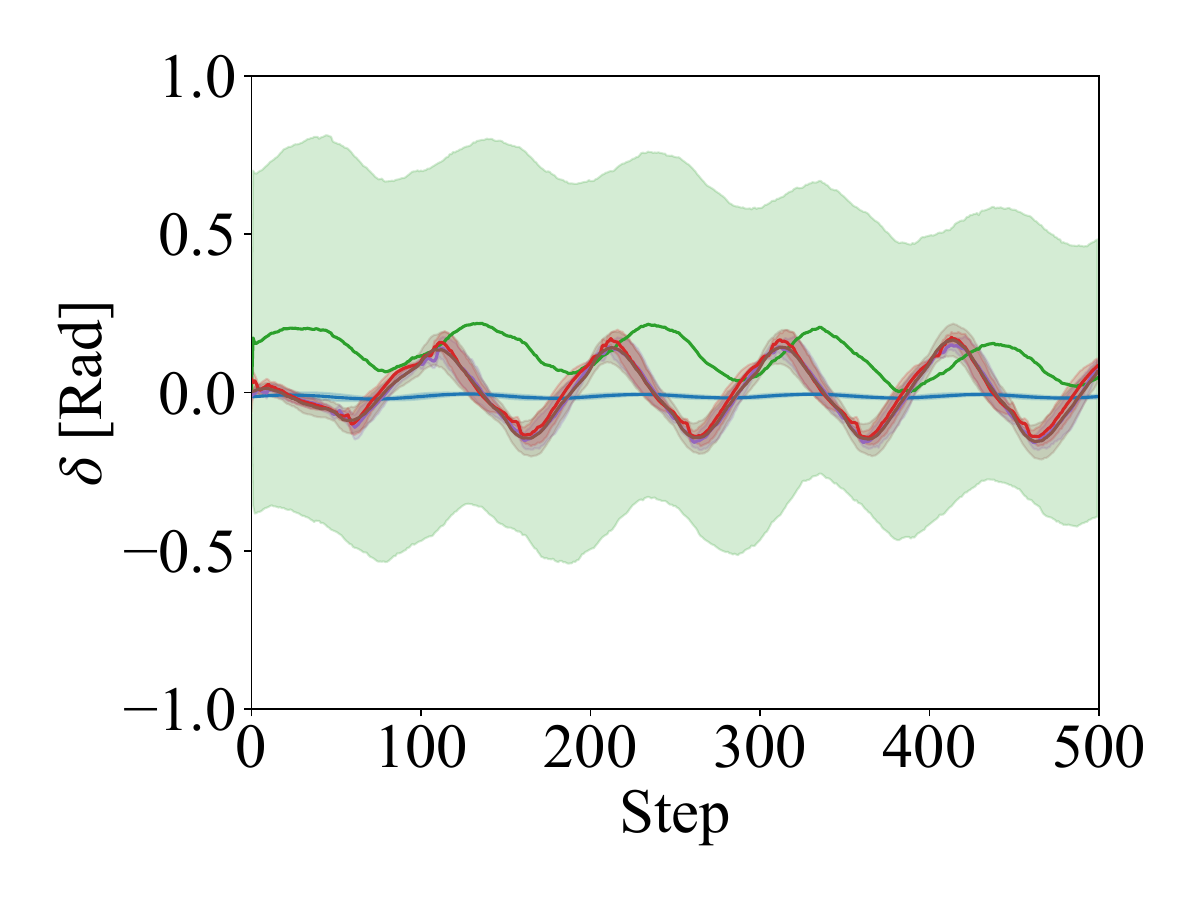}
        \label{sensor_net:2}
    }
    \\
    \vspace{-0.5cm}
    \subfloat{
        \includegraphics[width=0.225\textwidth]{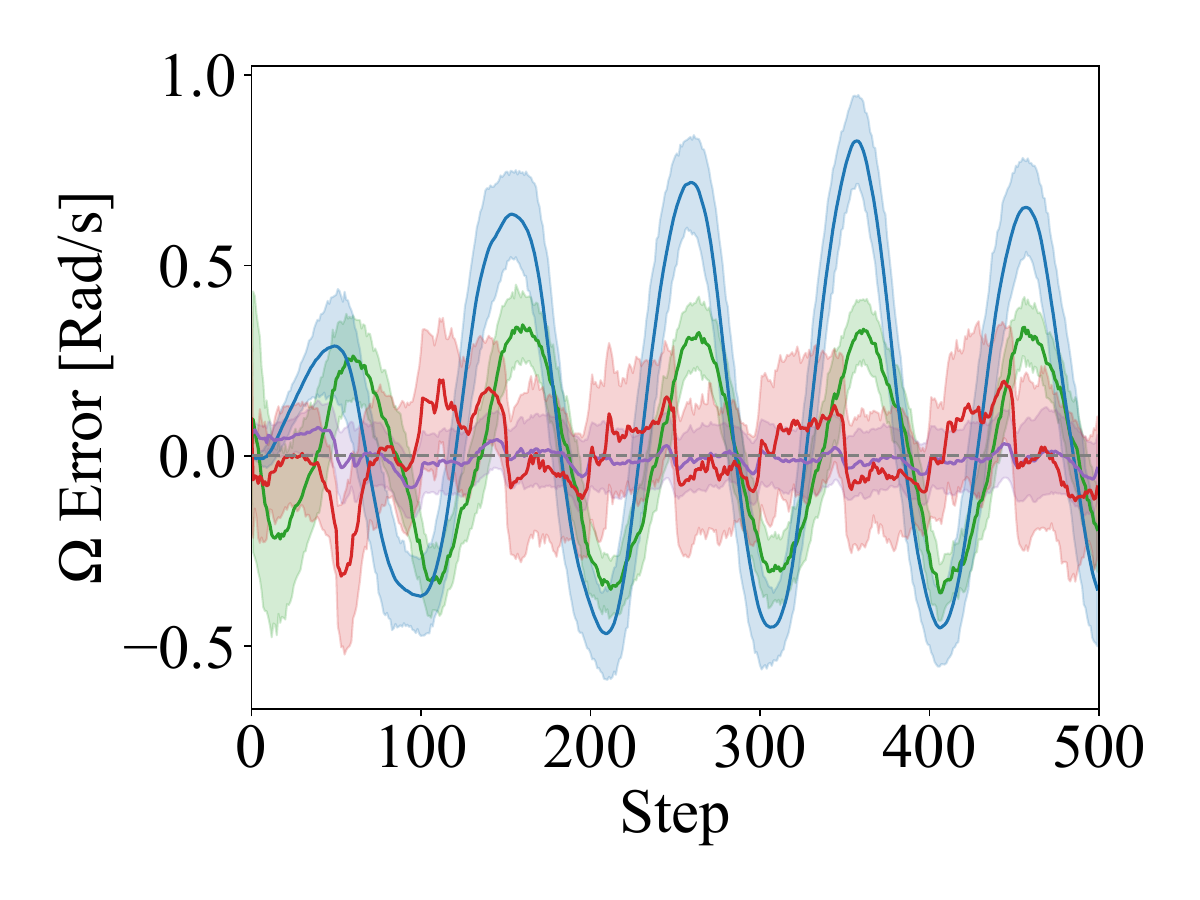}
        \label{sensor_net:3}
    }
    \hfill
    \subfloat{
        \includegraphics[width=0.225\textwidth]{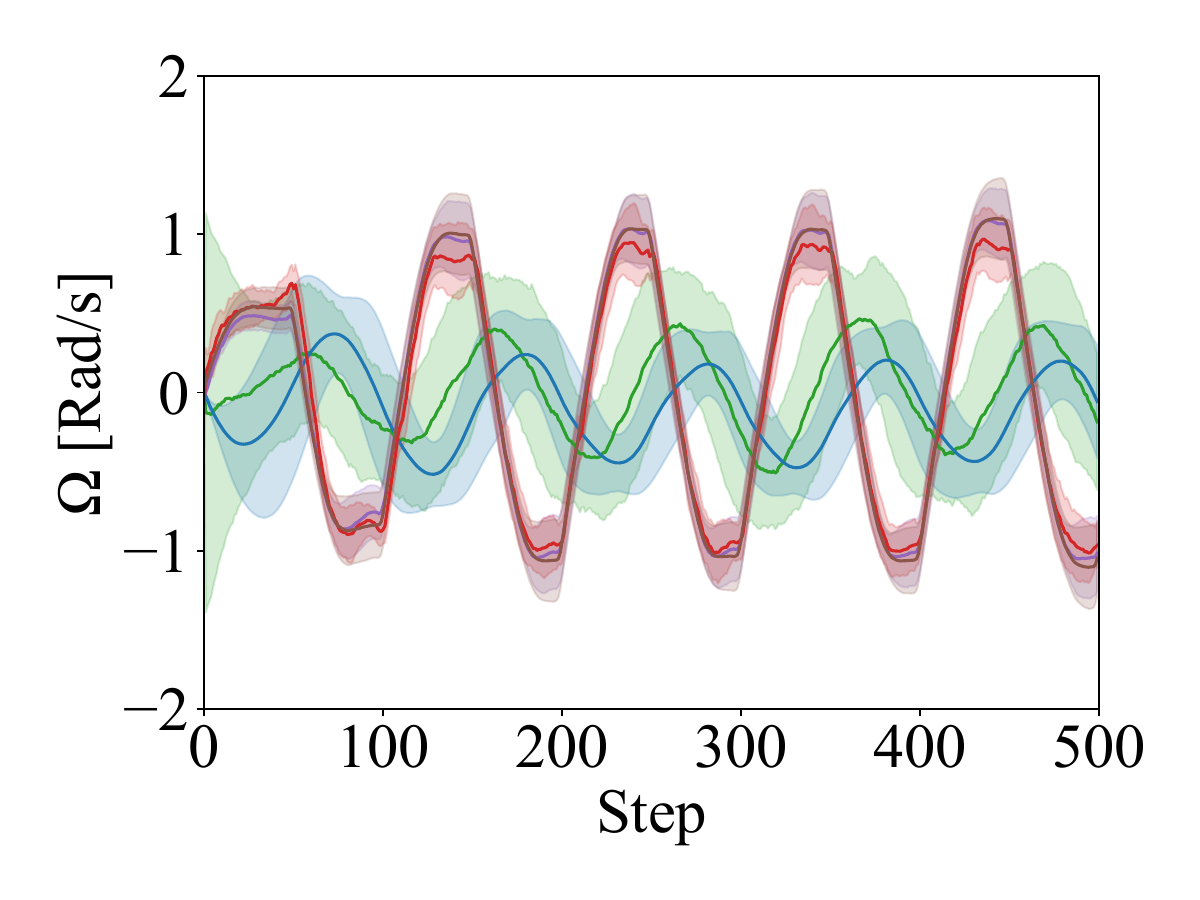}
        \label{sensor_net:4}
    }
    \\
    \vspace{-0.5cm}
    \subfloat{
        \includegraphics[width=0.35\textwidth]{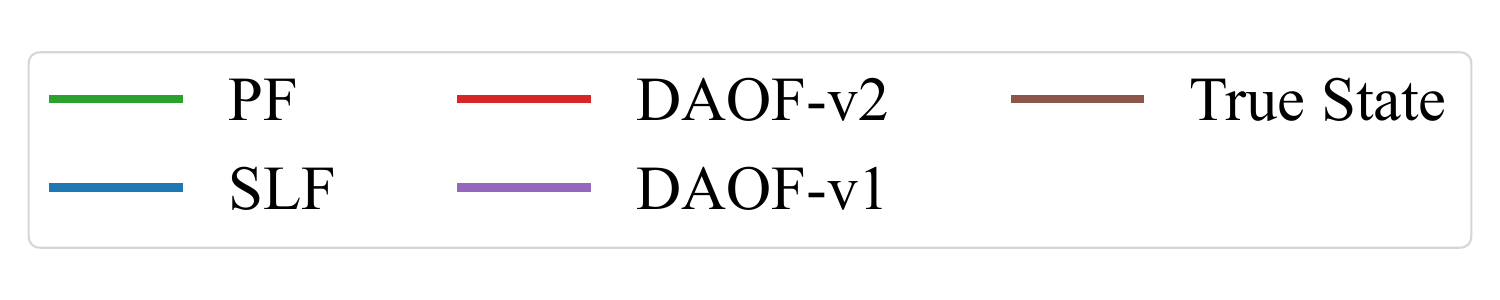}
        \label{sensor_net:legend}
        \phantomsection
    }
    \vspace{-0.3cm}
    \caption{Error and state plot for experiment I.}
    \label{fig.sensor_net_error}
\end{figure}

\begin{figure}[!h]
\centering
\includegraphics[width=0.45\textwidth]{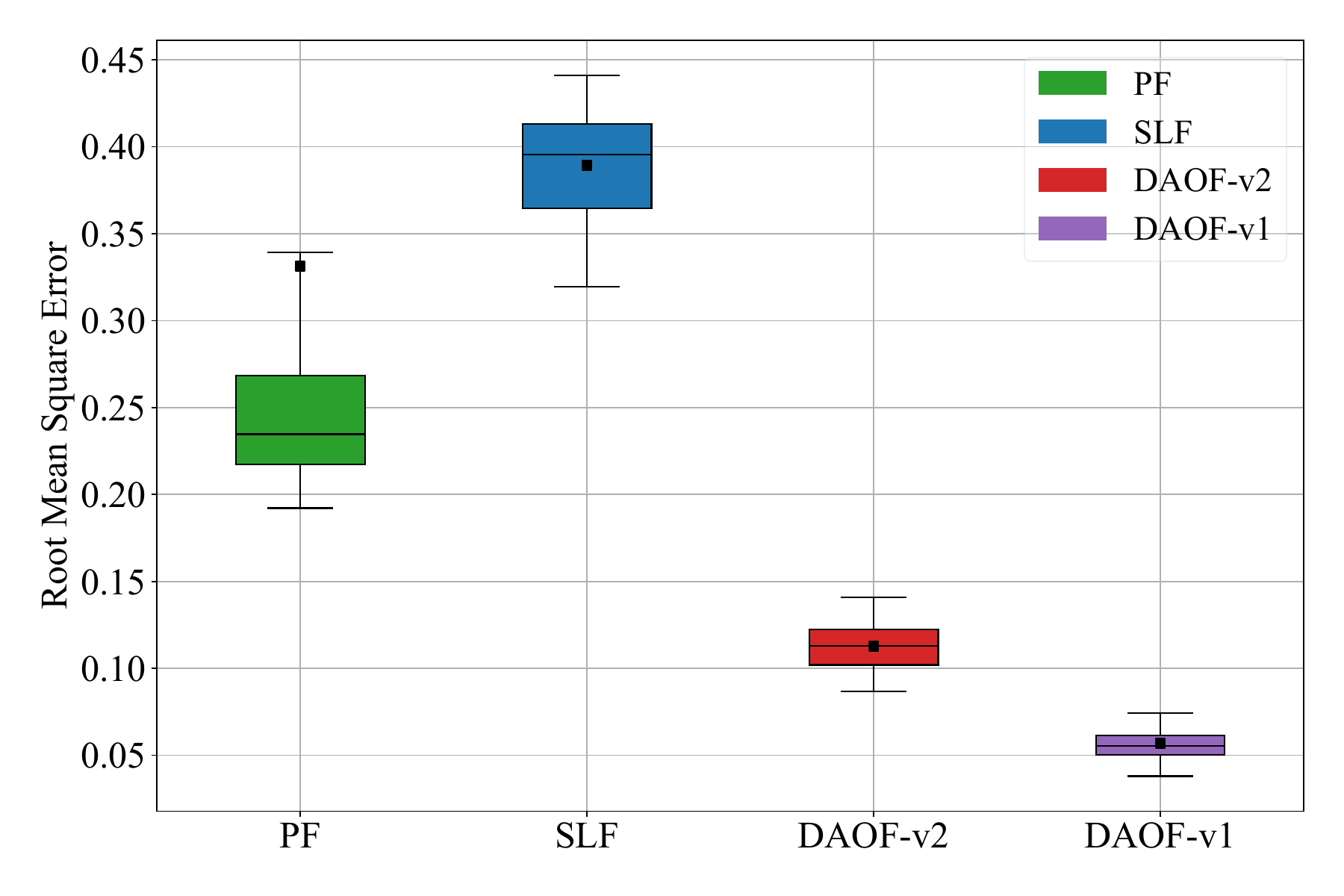}
\caption{Box plot of RMSE for PF, SLF, DAOF-v1, and DAOF-v2, for experiment I.}
\label{fig.sin_cos}
\end{figure}

\begin{figure}[!h]
    \centering
    \includegraphics[width=0.45\textwidth]{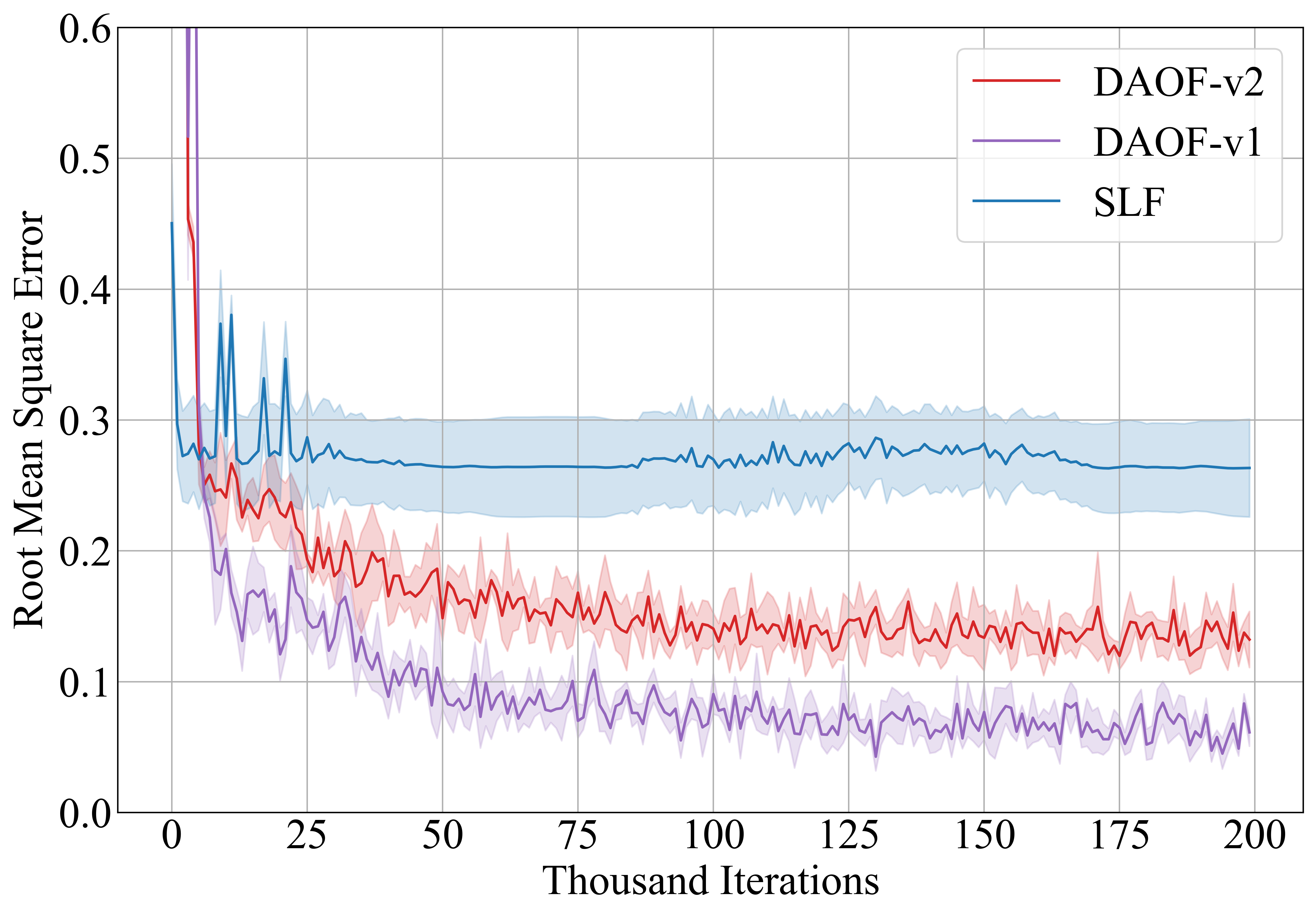}
    \caption{Comparison of training processes of SLF, DAOF-v1 and DAOF-v2 for experiment I.} 
    \label{fig:comparison_training}
\end{figure}

\begin{figure}[!h]
    \centering
    \includegraphics[width=0.45\textwidth]{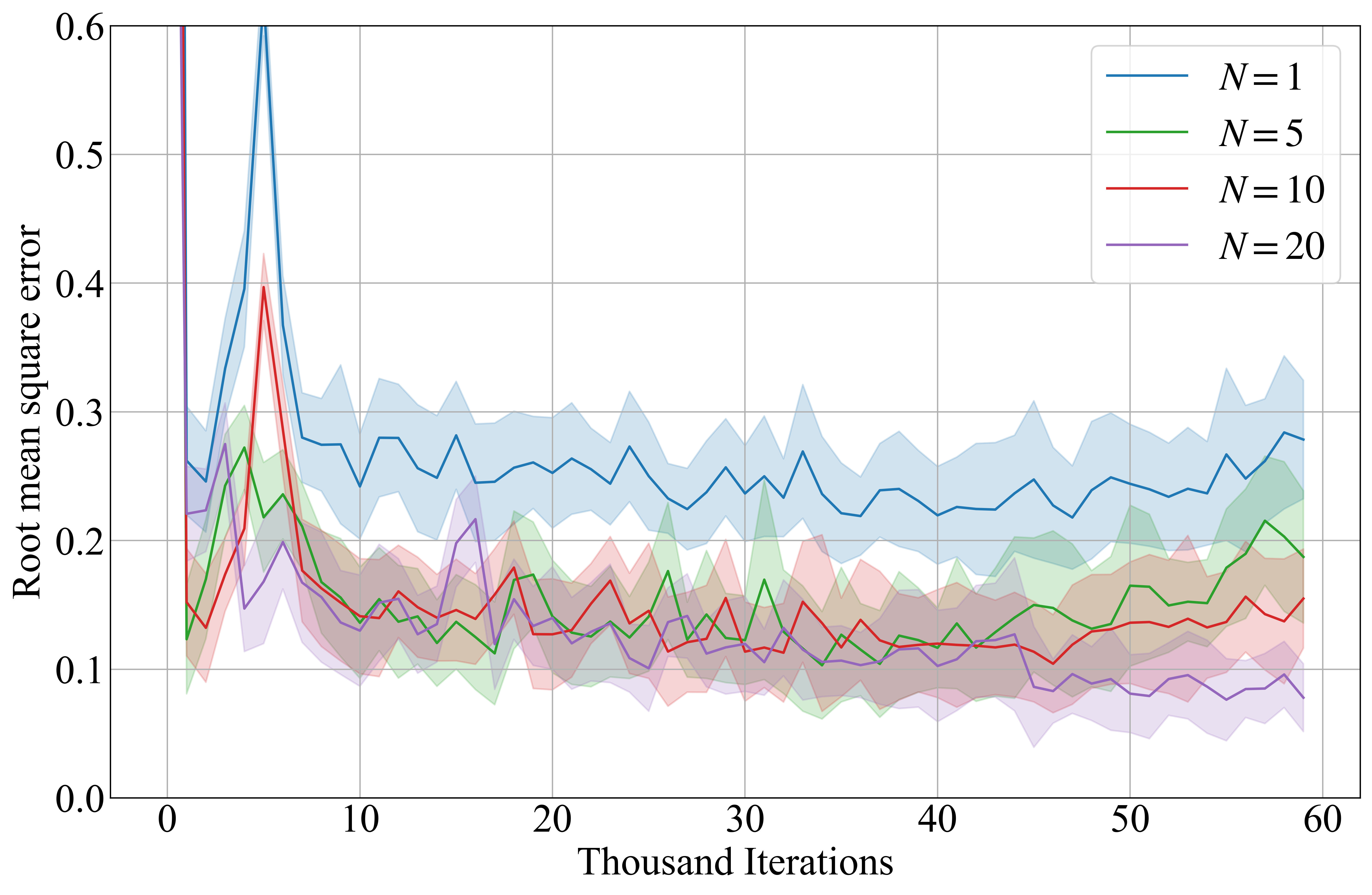}
    \caption{Comparison of different sliding window lengths.} 
    \label{fig:slidingwindows}
\end{figure}

The comparative results are detailed in Table \ref{tab:sideslipe_exp}, which shows that both structures of DAOF demonstrate superior estimation accuracy for both states compared to the baseline methods. DAOF-v1 achieves an impressively low RMSE of $\delta$ (0.03 rad) and $\Omega$ (0.09 rad/s), while maintaining a low computational cost (0.65 ms). As shown in Fig. \ref{fig.sensor_net_error}, compared to the baseline methods, the estimation errors of DAOF-v1 and DAOF-v2 for both states are closer to zero, and the estimated states are more aligned with the true states. Fig. \ref{fig.sin_cos} also demonstrates that DAOF-v1 exhibits the lowest average RMSE among the compared methods, with the least fluctuation.
As illustrated in Fig. \ref{fig:comparison_training}, both structures of DAOF require more iterations to converge compared to SLF, as two networks need to be trained for DAOF whereas only one is needed for SLF. However, DAOF achieves smaller RMSE upon convergence compared to SLF.
The enhanced estimation capabilities of our proposed method over SLF can be largely attributed to its actor-critic architecture. The PEV process employs recursive updates, where the value function estimate is updated through a weighted average of the current estimate and new observations. This approach smooths out noise in the value function estimation process, diminishing the impact of individual sample variability, and thereby reduces the variance of policy gradients \cite{sutton1999policy}. Consequently, the actor component can refine policies with a more stable basis, thereby improving the generalization capability of the policy. Furthermore, SLF tends to concentrate solely on minimizing the error associated with available samples, often overlooking the long-term consequences of policy decisions. This nearsighted optimization objective results in suboptimal performance, particularly in intricate and dynamic systems. In contrast, our approach is designed to address an infinite-horizon optimization challenge, which produces the current estimate by leveraging the Bellman equation, thereby focusing on formulating policies that minimize long-term costs. As a result, our training algorithm yields more holistic and optimal policies that are better suited for complexities of dynamic systems. As illustrated in Fig \ref{fig:slidingwindows}, we we investigate the impact of different sliding window lengths on the training process and performance of DAOF-v1. The results indicate that an increased sliding window length correlates with a reduction in RMSE after training convergence, while the computational cost increases. This reflects the trade-off between accuracy and computational efficiency that need to be carefully considered when determining the sliding window length.

\begin{table}[!h]
    \renewcommand{\arraystretch}{1.3}
    \caption{Performance on Experiment I}
    \label{tab:sideslipe_exp}
    \centering
    \begin{tabular}{cccc}
        \hline\hline
                  & \multicolumn{2}{c}{RMSE} & \multirow{2}{*}{\begin{tabular}[c]{@{}c@{}}Computational\\ Cost (ms)\end{tabular}} \\ \cline{2-3}
                  & $\delta$ [rad] & $\Omega$ [rad/s] &                                                                                   \\ \hline
        UKF       & 39.91 & 0.94    & 0.28   \\
        PF        & 0.19 & 0.47    & 21.4   \\ 
        SLF       & 0.14 & 0.68    & 0.92   \\ 
        DAOF-v2      & 0.04 & 0.19   & 0.63   \\
        DAOF-v1     & 0.03 & 0.09   & 0.65   \\\hline\hline
        \end{tabular}
    \vspace{-0.5cm}
\end{table}

\begin{table}[!h]
    \renewcommand{\arraystretch}{1.3}
    \caption{Performance for different sliding window lengths}
    \label{tab:sideslipe_exp}
    \centering
    \begin{tabular}{ccc}
        \hline\hline
                  Length of Sliding Window
                  &RMSE 
                  &Computational Cost (ms)                                                                                   \\ \hline
        $N=1$       & 0.29 & 0.29   \\
        $N=5$        & 0.21    & 0.29   \\ 
        $N=10$       & 0.16   & 0.31   \\ 
        $N=20$      & 0.11   & 0.34    \\\hline\hline
        \end{tabular}
\end{table}

\subsection{Experiment II: Estimation of a 14-DOF Vehicle Model}
\begin{figure}[!h]
    \centering
    \subfloat{
        \includegraphics[width=0.225\textwidth]{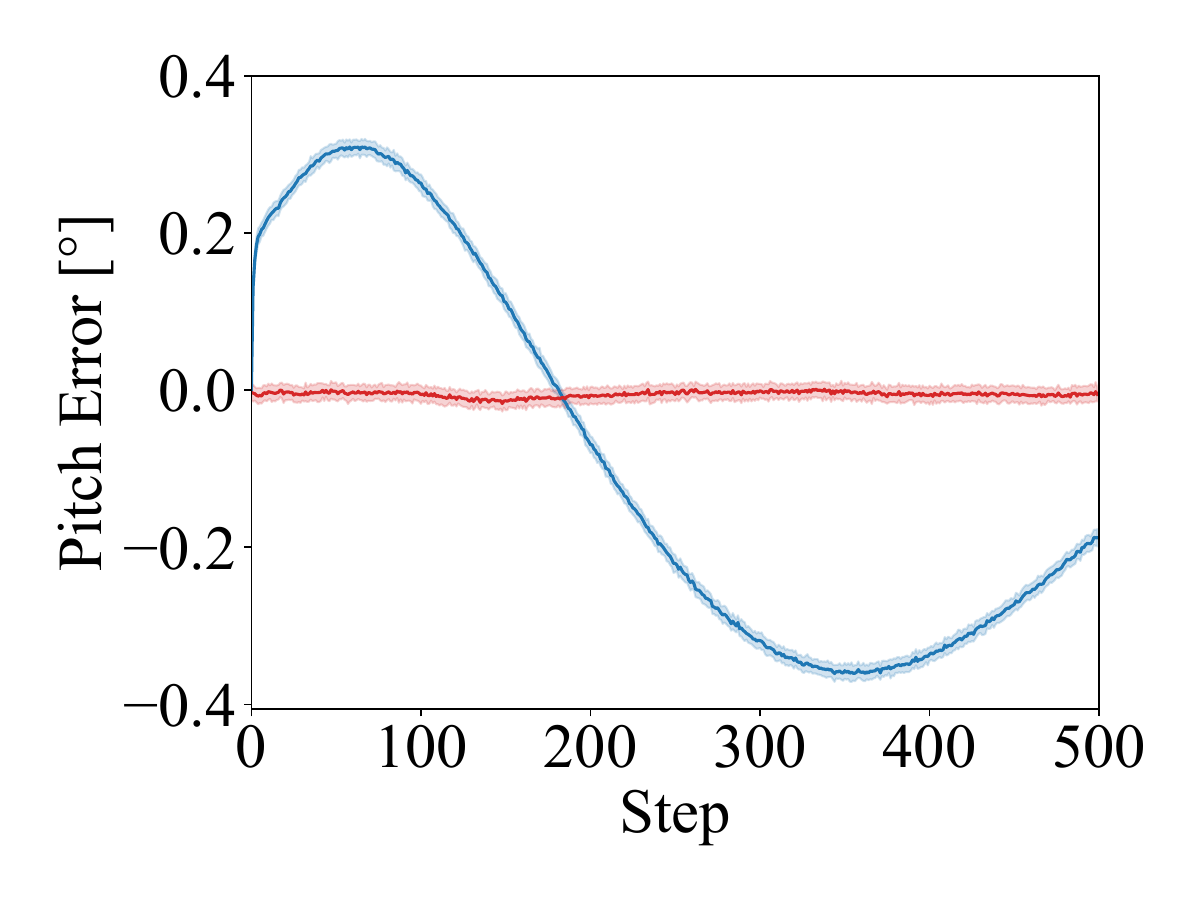}
    }
    \hfill
    \subfloat{
        \includegraphics[width=0.225\textwidth]{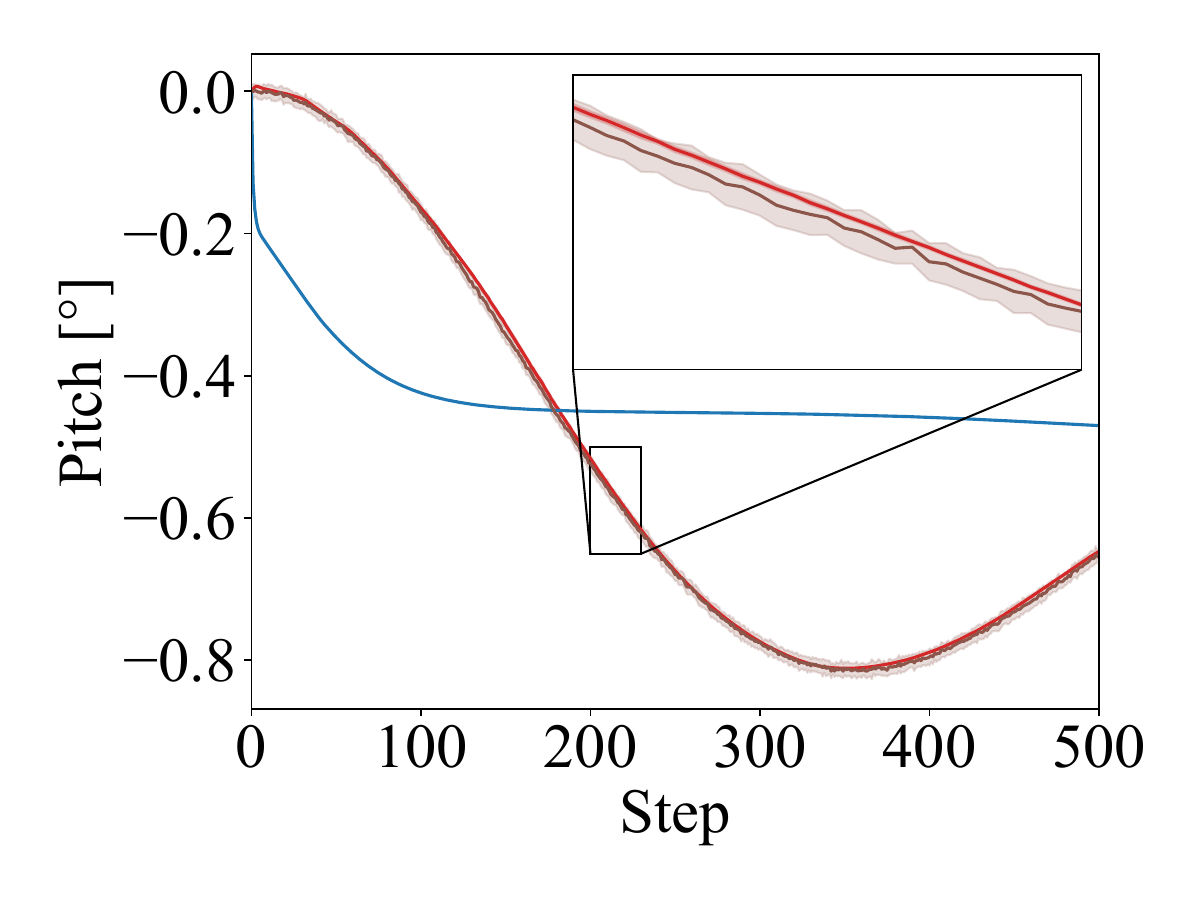}
    }
    \\
    \vspace{-0.5cm}
    \subfloat{
        \includegraphics[width=0.225\textwidth]{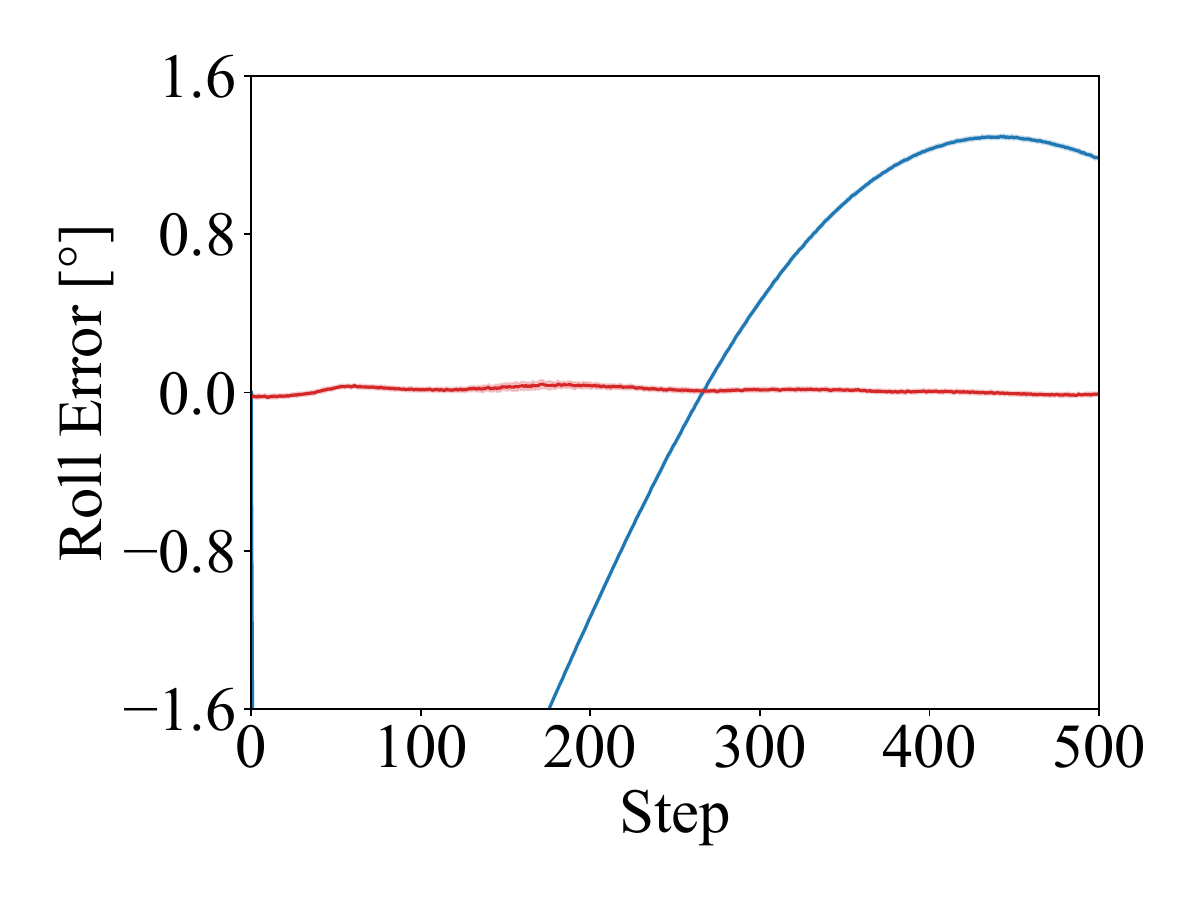}
    }
    \hfill
    \subfloat{
        \includegraphics[width=0.225\textwidth]{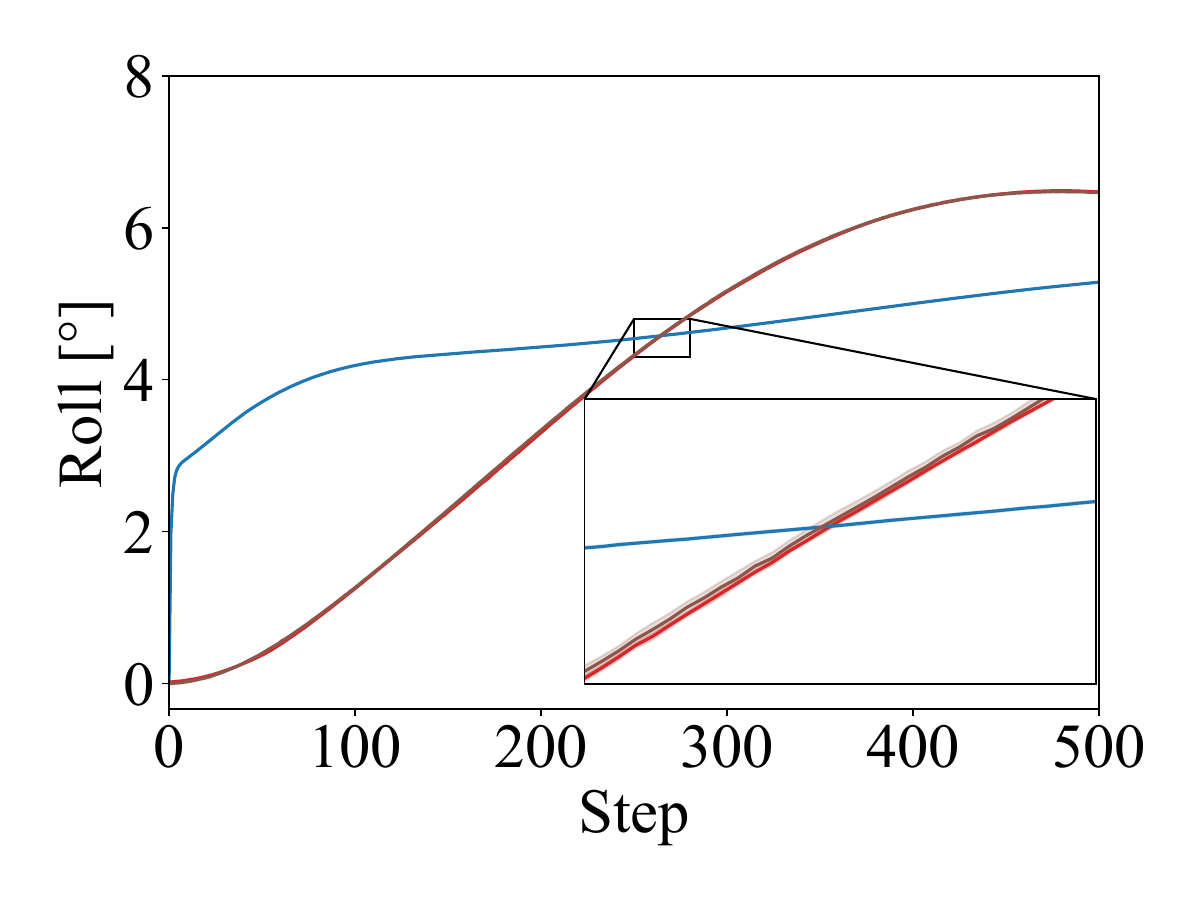}
    }
    \\
    \vspace{-0.5cm}
    \subfloat{
        \includegraphics[width=0.225\textwidth]{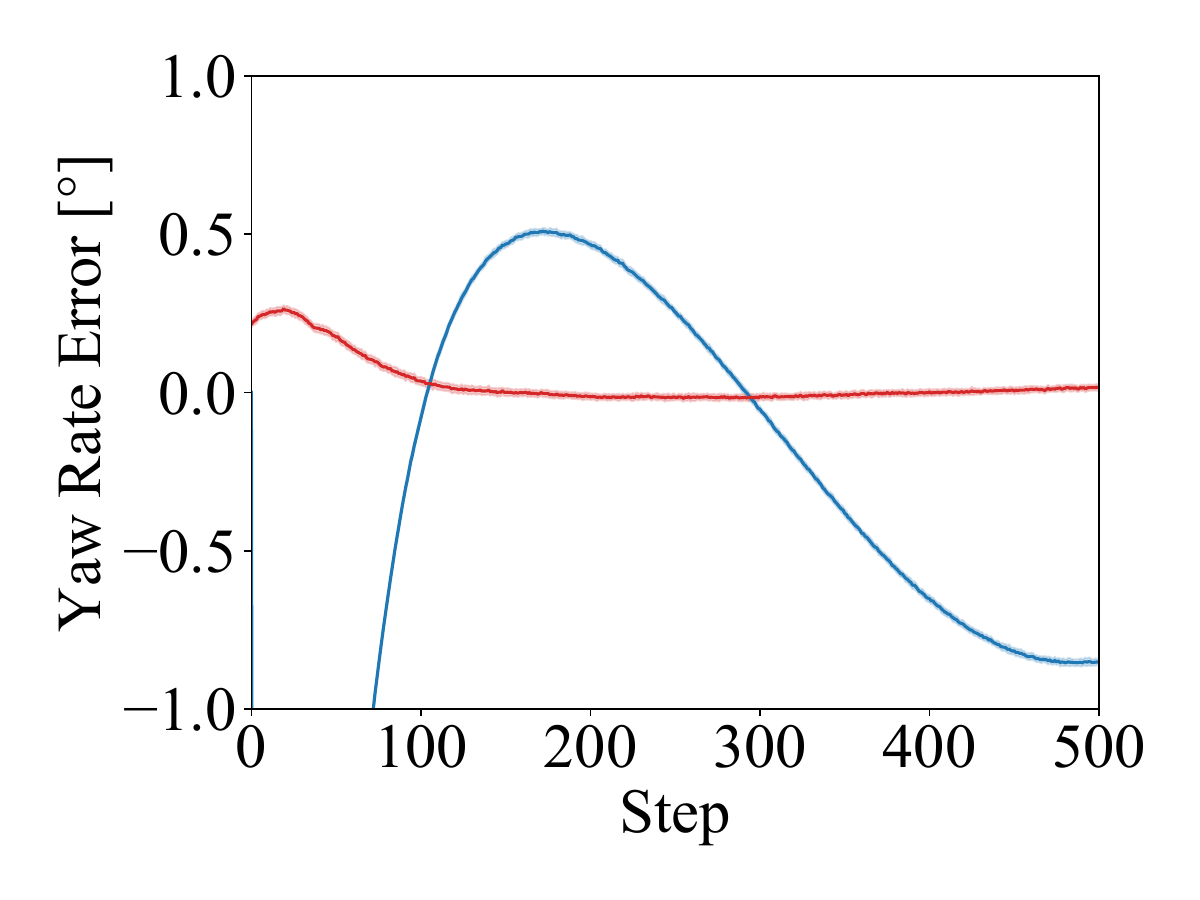}
    }
    \hfill
    \subfloat{
        \includegraphics[width=0.225\textwidth]{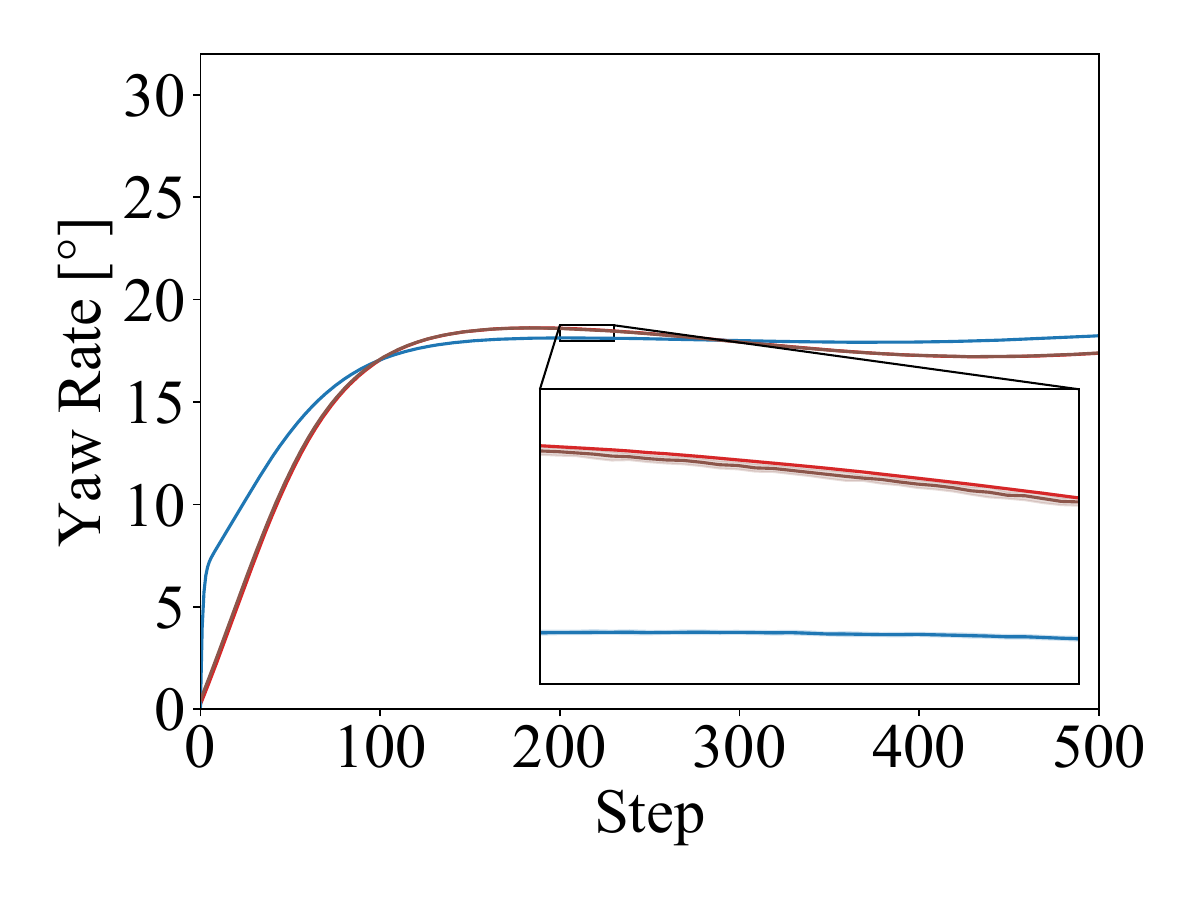}
    }
    \\
    \vspace{-0.5cm}
    \subfloat{
        \includegraphics[width=0.225\textwidth]{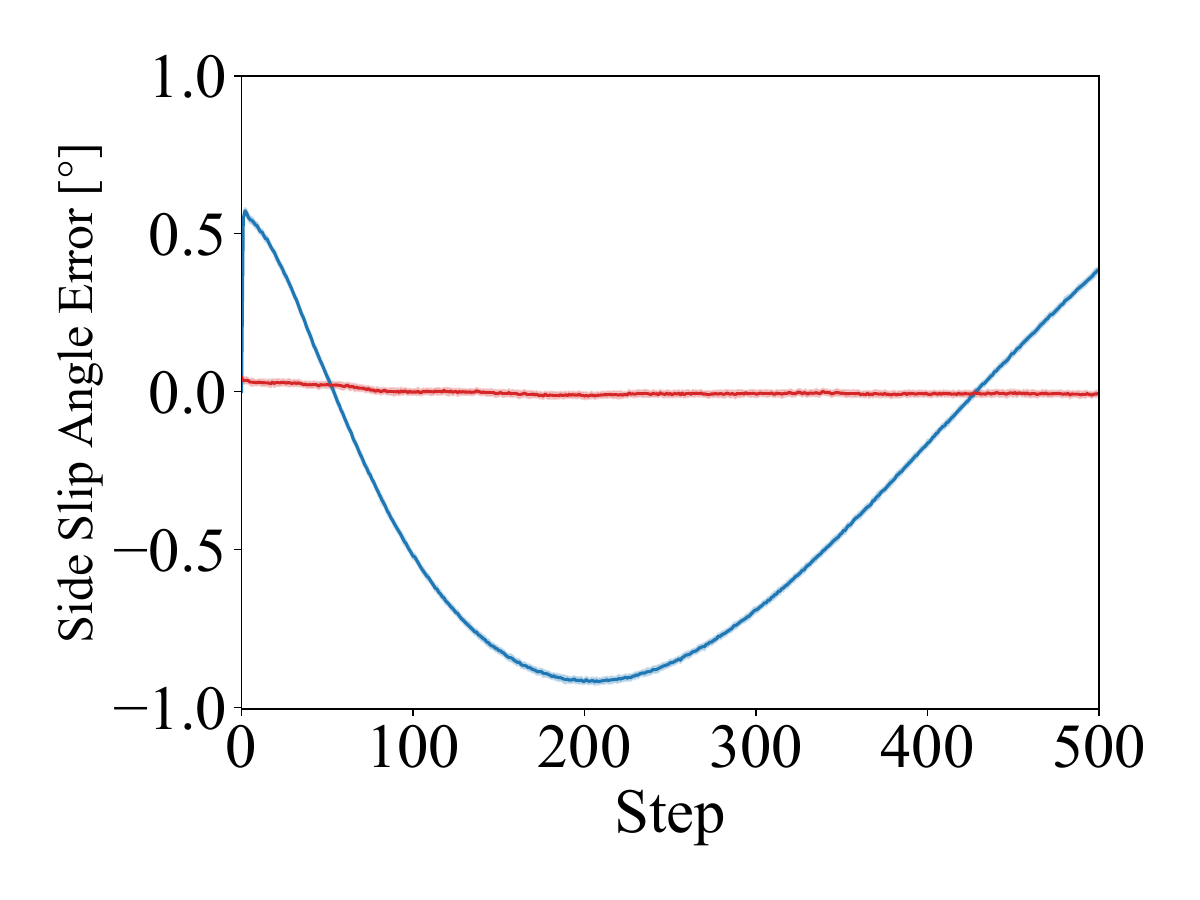}
    }
    \hfill
    \subfloat{
        \includegraphics[width=0.225\textwidth]{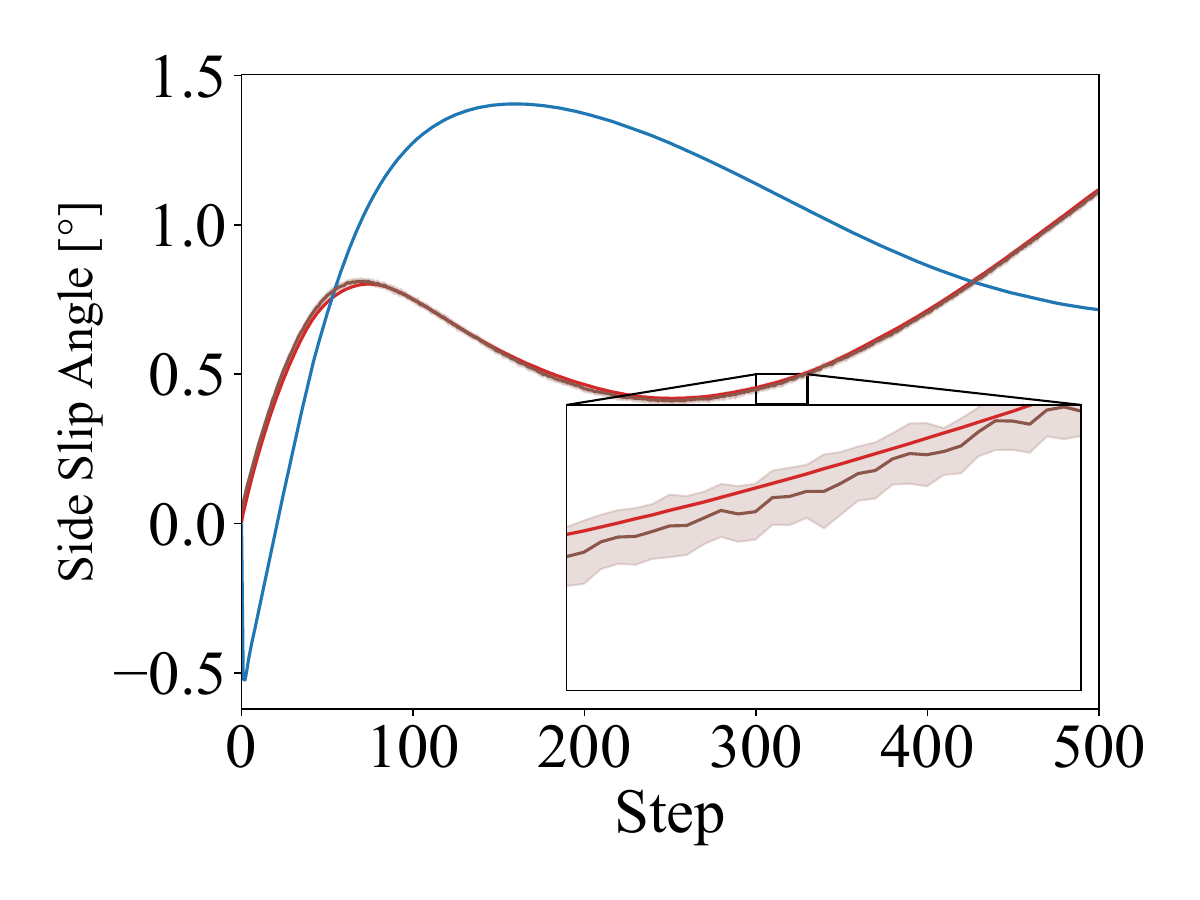}
    }
    \\
    \vspace{-0.5cm}
    \subfloat{
        \includegraphics[width=0.225\textwidth]{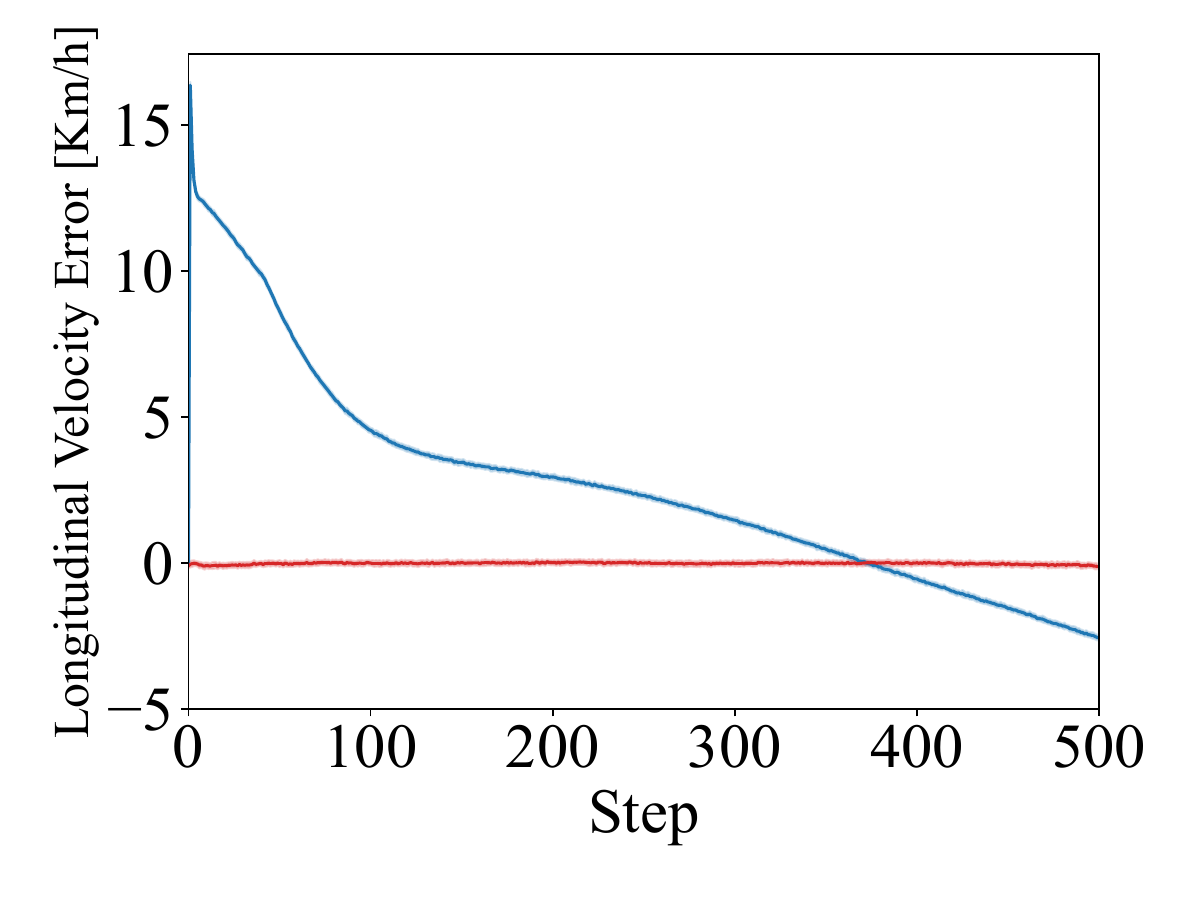}
    }
    \hfill
    \subfloat{
        \includegraphics[width=0.225\textwidth]{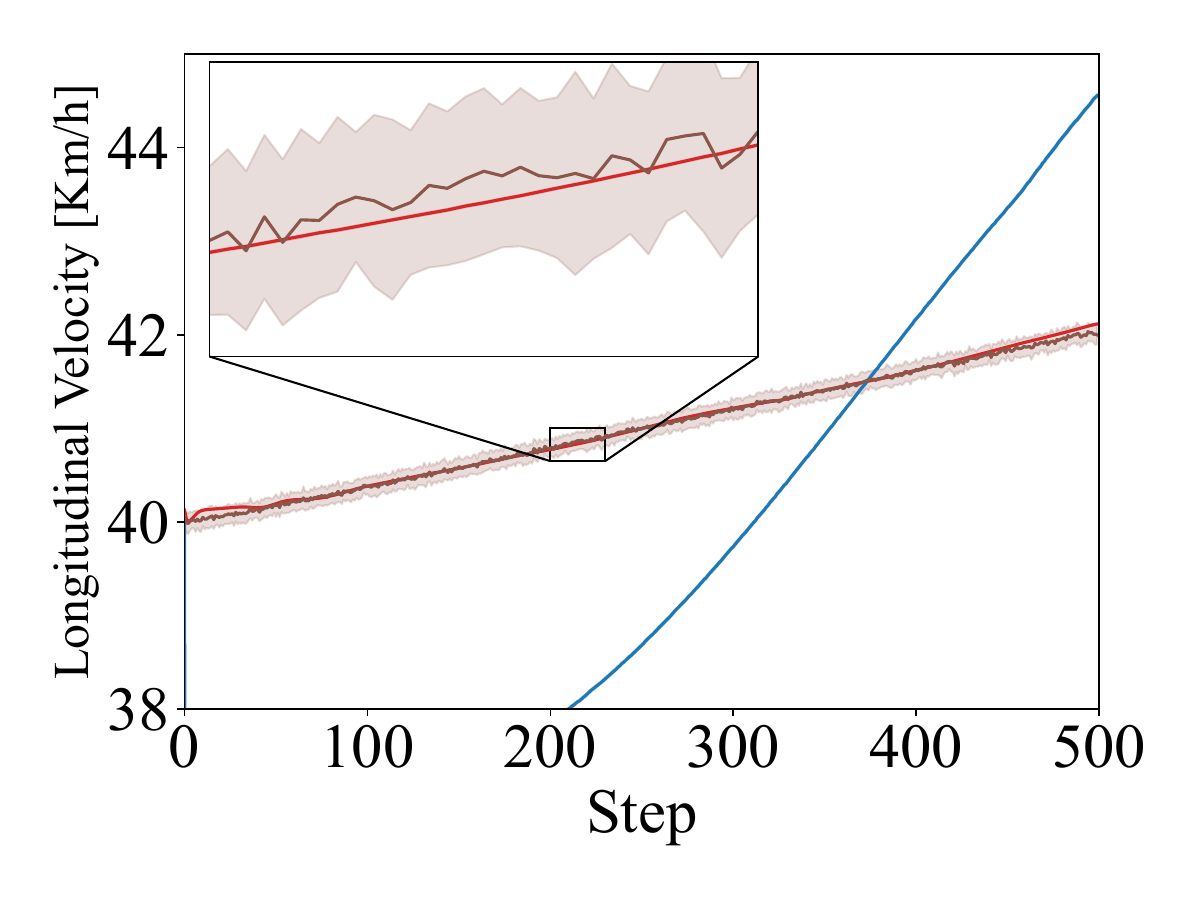}
    }
    \\
    \vspace{-0.5cm}
    \subfloat{
        \includegraphics[width=0.225\textwidth]{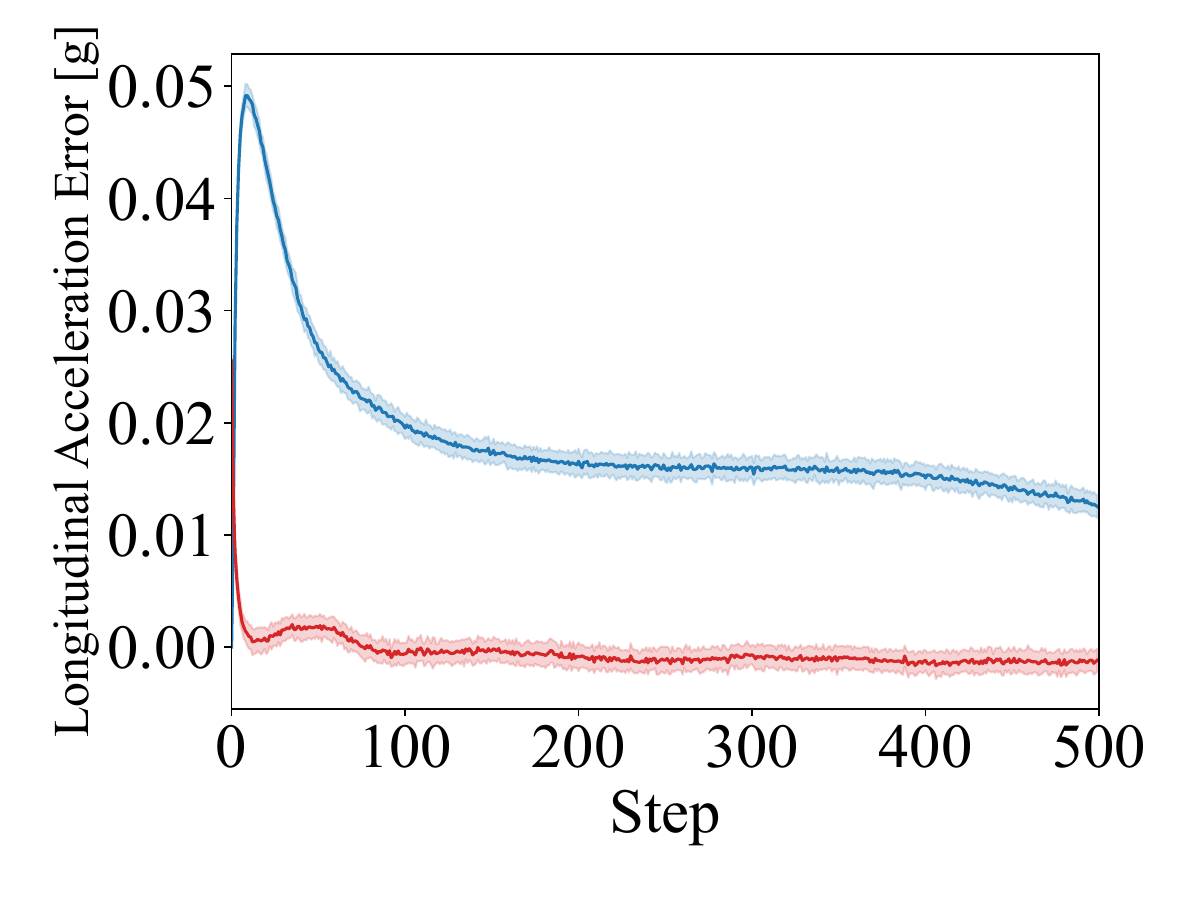}
    }
    \hfill
    \subfloat{
        \includegraphics[width=0.225\textwidth]{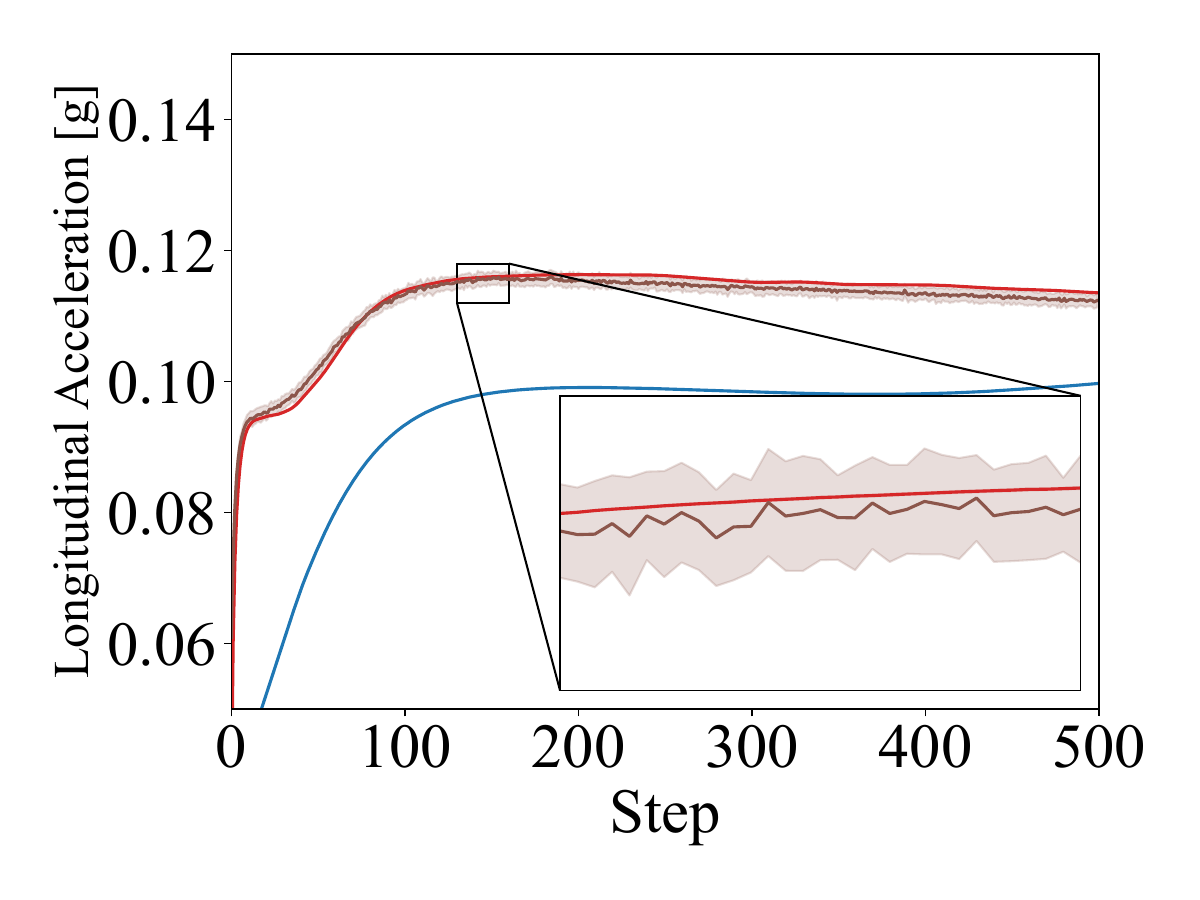}
    }
    \\
    \vspace{-0.5cm}
    \subfloat{
        \includegraphics[width=0.35\textwidth]{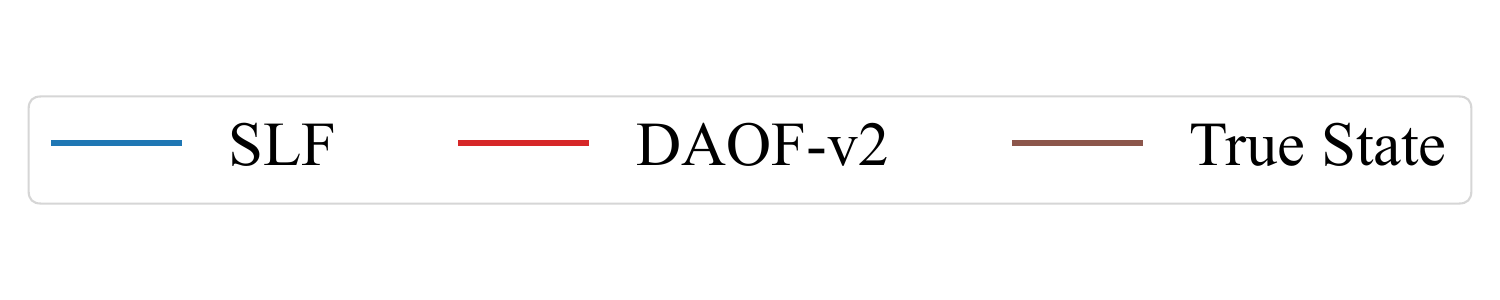}
        \phantomsection
    }
    \vspace{-0.5cm}
    \caption{Error and state plot for experiment II.}
    \label{fig.experiment II}
\end{figure}

We employ a 14-DOF vehicle dynamics model to achieve the most accurate possible representation of vehicle states \cite{li2023maximizing}. This model encompasses a 6-DOF vehicle body module, a 4-DOF wheel module and a 4-DOF suspension module. By analyzing the dynamic and kinematic properties of these three modules, the differential equations of each module can be derived and then combined to obtain the integrated 14-DOF model. 
\begin{table}[!h]
    \renewcommand{\arraystretch}{1.3}
    \setlength{\tabcolsep}{1pt}
    \caption{Performance on Experiment II}
    \label{tab:delta}
    \centering
    \begin{tabular}{cccccccc}
        \hline\hline
                  & \multicolumn{6}{c}{RMSE} & \multirow{2}{*}{\begin{tabular}[c]{@{}c@{}}Computational\\ Cost (ms)\end{tabular}} \\ \cline{2-7}
                  & $\theta$ [$^{\circ}$] & $\phi$ [$^{\circ}$] & $\dot{\varphi}$ [$^{\circ}/s$] & $\beta$ [$^{\circ}$] &$V_{x}$ [Km/h] & $a_{x}$ [g]                                                                                  \\ \hline
        SLF & 0.262 & 1.863 & 1.513 & 0.587 & 4.545 & 0.020 & 0.45\\ 
        DAOF-v2  & 0.012 & 0.022 & 0.082& 0.015 & 0.105 & 0.002 & 0.63\\
        \hline\hline
        \end{tabular}
\end{table}
Considering the complexity and mutual coupling of the differential equations, the conventional Euler solving method encounters low accuracy. Therefore, we use the fourth-order Runge-Kutta method to ensure both accuracy and computational efficiency. However, using this method makes it impractical to explicitly provide the transition and observation functions. Consequently, we treat this vehicle dynamics model as a non-interactive data source and perform state estimation solely based on the data generated during operation of the model. This approach requires model-free filtering capability, which only DAOF-v2 and SLF can provide. In contrast, typical Bayesian filters are all ineffective. We evaluate the performance of DAOF-v2 in a large curvature turning condition. The states we need to estimate involve the pitch angle, longitudinal velocity and acceleration, yaw rate, roll angle and side slip angle. To simulate sensor observations in a real vehicle, we use the following data as observation inputs: rotational speed of each wheel for wheel speed sensor, steering wheel angle for steering angle sensor, longitudinal, lateral, and vertical accelerations, as well as angular velocity for Inertial Measurement Unit (IMU), and longitudinal and lateral position and velocity for Global Navigation Satellite System (GNSS). Detailed data in Table \ref{tab:delta} shows that while SLF fails to estimate several states, DAOF-v2 achieves low RMSE and computational cost. As shown in Fig. \ref{fig.experiment II}, the estimation errors of DAOF-v2 for all states are close to zero, and
the estimated states are well aligned with the true states.

\begin{table}[!h]
\centering
\caption{Detailed Parameters for Training}
\begin{tabular}{ll}
\hline
\textbf{Training parameters}              & \textbf{Value}    \\
\hline
\textbf{Shared}\\
\quad Optimizer                        & ADAM ($\beta_1$=0.9, $\beta_2$=0.999) \\
\quad Actor learning rate              & 1e-4\\
\quad Critic learning rate             & 1e-4\\
\quad Discount factor $(\gamma)$       & 0.99\\
\quad Policy update interval           & 2\\
\quad Target smoothing coefficient $\tau$     & 0.005\\
\quad Reward scale                     & 1\\
\quad Number of hidden layers          & 3        \\
\quad Number of hidden units per layer & 256      \\
\quad Replay buffer size               & $1\times10^6$ \\
\quad Sample batch size                & 20 \\
\hline
\textbf{ValueNet}                         &          \\
\quad Approximation function           & MLP      \\
\quad Activation function of hidden layer     & Gelu      \\
\quad Activation function of output layer     & Identity \\
\hline
\textbf{PolicyNet}                        &          \\
\quad Approximation Function           & MLP      \\
\quad Activation function of hidden layer     & Gelu     \\
\quad Activation function of output layer     & Identity   \\
\hline
\end{tabular}
\label{table: training configuration}
\end{table}

\section{Conclusion}

In this work, we formulate a filtering problem, termed MFP, and bridge it with RL. To solve MFP, we derive DAOF, a filter tailored for nonlinear and non-Gaussian systems. We design two structures for DAOF to address diverse filtering scenarios, depending on the presence or absence of an explicit system model. Using an actor-critic framework, we apply an RL algorithm to train the policy for DAOF. Experimental results demonstrate the superior accuracy and computational efficiency of DAOF-v1 compared to conventional nonlinear filters. Furthermore, DAOF-v2 demonstrates the ability to tackle filtering problems in the absence of an explicit system model, enabling direct estimation of complex real-world systems without the need for model simplifications or strict assumptions.




\bibliographystyle{IEEEtran}
\bibliography{ref}

\begin{thebibliography}{10}
\providecommand{\url}[1]{#1}
\csname url@samestyle\endcsname
\providecommand{\newblock}{\relax}
\providecommand{\bibinfo}[2]{#2}
\providecommand{\BIBentrySTDinterwordspacing}{\spaceskip=0pt\relax}
\providecommand{\BIBentryALTinterwordstretchfactor}{4}
\providecommand{\BIBentryALTinterwordspacing}{\spaceskip=\fontdimen2\font plus
\BIBentryALTinterwordstretchfactor\fontdimen3\font minus \fontdimen4\font\relax}
\providecommand{\BIBforeignlanguage}[2]{{%
\expandafter\ifx\csname l@#1\endcsname\relax
\typeout{** WARNING: IEEEtran.bst: No hyphenation pattern has been}%
\typeout{** loaded for the language `#1'. Using the pattern for}%
\typeout{** the default language instead.}%
\else
\language=\csname l@#1\endcsname
\fi
#2}}
\providecommand{\BIBdecl}{\relax}
\BIBdecl

\bibitem{sarkka2023bayesian}
S.~S{\"a}rkk{\"a} and L.~Svensson, \emph{Bayesian filtering and smoothing}.\hskip 1em plus 0.5em minus 0.4em\relax Cambridge university press, 2023, vol.~17.

\bibitem{kalman1960new}
R.~E. Kalman, ``A new approach to linear filtering and prediction problems,'' 1960.

\bibitem{smith1962application}
G.~L. Smith, S.~F. Schmidt, and L.~A. McGee, \emph{Application of statistical filter theory to the optimal estimation of position and velocity on board a circumlunar vehicle}.\hskip 1em plus 0.5em minus 0.4em\relax National Aeronautics and Space Administration, 1962, vol. 135.

\bibitem{julier1997new}
S.~J. Julier and J.~K. Uhlmann, ``New extension of the kalman filter to nonlinear systems,'' in \emph{Signal processing, sensor fusion, and target recognition VI}, vol. 3068.\hskip 1em plus 0.5em minus 0.4em\relax Spie, 1997, pp. 182--193.

\bibitem{perea2007nonlinearity}
L.~Perea, J.~How, L.~Breger, and P.~Elosegui, ``Nonlinearity in sensor fusion: divergence issues in ekf, modified truncated gsf, and ukf,'' in \emph{AIAA Guidance, Navigation and Control Conference and Exhibit}, 2007, p. 6514.

\bibitem{liu1998sequential}
J.~S. Liu and R.~Chen, ``Sequential monte carlo methods for dynamic systems,'' \emph{Journal of the American statistical association}, vol.~93, no. 443, pp. 1032--1044, 1998.

\bibitem{rao2001constrained}
C.~V. Rao, J.~B. Rawlings, and J.~H. Lee, ``Constrained linear state estimation—a moving horizon approach,'' \emph{Automatica}, vol.~37, no.~10, pp. 1619--1628, 2001.

\bibitem{al2019deep}
M.~K. Al-Sharman, Y.~Zweiri, M.~A.~K. Jaradat, R.~Al-Husari, D.~Gan, and L.~D. Seneviratne, ``Deep-learning-based neural network training for state estimation enhancement: Application to attitude estimation,'' \emph{IEEE Transactions on Instrumentation and Measurement}, vol.~69, no.~1, pp. 24--34, 2019.

\bibitem{jin2021new}
X.-B. Jin, R.~J. Robert~Jeremiah, T.-L. Su, Y.-T. Bai, and J.-L. Kong, ``The new trend of state estimation: From model-driven to hybrid-driven methods,'' \emph{Sensors}, vol.~21, no.~6, p. 2085, 2021.

\bibitem{bai2023state}
Y.~Bai, B.~Yan, C.~Zhou, T.~Su, and X.~Jin, ``State of art on state estimation: Kalman filter driven by machine learning,'' \emph{Annual Reviews in Control}, vol.~56, p. 100909, 2023.

\bibitem{revach2022kalmannet}
G.~Revach, N.~Shlezinger, X.~Ni, A.~L. Escoriza, R.~J. Van~Sloun, and Y.~C. Eldar, ``Kalmannet: Neural network aided kalman filtering for partially known dynamics,'' \emph{IEEE Transactions on Signal Processing}, vol.~70, pp. 1532--1547, 2022.

\bibitem{chen2021dynanet}
C.~Chen, C.~X. Lu, B.~Wang, N.~Trigoni, and A.~Markham, ``Dynanet: Neural kalman dynamical model for motion estimation and prediction,'' \emph{IEEE Transactions on Neural Networks and Learning Systems}, vol.~32, no.~12, pp. 5479--5491, 2021.

\bibitem{karl2017deep}
M.~Karl, M.~Soelch, J.~Bayer, and P.~van~der Smagt, ``Deep variational bayes filters: Unsupervised learning of state space models from raw data,'' in \emph{International Conference on Learning Representations}, 2017.

\bibitem{allan2019moving}
D.~A. Allan and J.~B. Rawlings, ``Moving horizon estimation,'' \emph{Handbook of model predictive control}, pp. 99--124, 2019.

\bibitem{li2023maximizing}
H.~Li, K.~Liu, B.~Zhao, N.~Xu, Y.~Huang, and Y.~Yin, ``Maximizing the effective quasi-usage rate for 4wimd-evs under combined-slip conditions,'' \emph{IEEE Transactions on Vehicular Technology}, 2023.

\bibitem{shakya2023reinforcement}
A.~K. Shakya, G.~Pillai, and S.~Chakrabarty, ``Reinforcement learning algorithms: A brief survey,'' \emph{Expert Systems with Applications}, vol. 231, p. 120495, 2023.

\bibitem{aliramezani2022modeling}
M.~Aliramezani, C.~R. Koch, and M.~Shahbakhti, ``Modeling, diagnostics, optimization, and control of internal combustion engines via modern machine learning techniques: A review and future directions,'' \emph{Progress in Energy and Combustion Science}, vol.~88, p. 100967, 2022.

\bibitem{arulkumaran2017deep}
K.~Arulkumaran, M.~P. Deisenroth, M.~Brundage, and A.~A. Bharath, ``Deep reinforcement learning: A brief survey,'' \emph{IEEE Signal Processing Magazine}, vol.~34, no.~6, pp. 26--38, 2017.

\bibitem{wang2022deep}
X.~Wang, S.~Wang, X.~Liang, D.~Zhao, J.~Huang, X.~Xu, B.~Dai, and Q.~Miao, ``Deep reinforcement learning: A survey,'' \emph{IEEE Transactions on Neural Networks and Learning Systems}, vol.~35, no.~4, pp. 5064--5078, 2022.

\bibitem{RN10}
J.~Morimoto and K.~Doya, ``Reinforcement learning state estimator,'' \emph{Neural computation}, vol.~19, no.~3, pp. 730--756, 2007.

\bibitem{cao2021reinforced}
W.~Cao, J.~Chen, J.~Duan, S.~E. Li, Y.~Lyu, Z.~Gu, and Y.~Zhang, ``Reinforced optimal estimator,'' \emph{IFAC-PapersOnLine}, vol.~54, no.~20, pp. 366--373, 2021.

\bibitem{anderson2012optimal}
B.~D. Anderson and J.~B. Moore, \emph{Optimal filtering}.\hskip 1em plus 0.5em minus 0.4em\relax Courier Corporation, 2012.

\bibitem{bellman1966dynamic}
R.~Bellman, ``Dynamic programming,'' \emph{science}, vol. 153, no. 3731, pp. 34--37, 1966.

\bibitem{li2023reinforcement}
S.~E. Li, \emph{Reinforcement learning for sequential decision and optimal control}.\hskip 1em plus 0.5em minus 0.4em\relax Springer, 2023.

\bibitem{duan2023dsact}
J.~Duan, W.~Wang, L.~Xiao, J.~Gao, and S.~E. Li, ``Dsac-t: Distributional soft actor-critic with three refinements,'' 2023.

\bibitem{pfeifer2019expectation}
T.~Pfeifer and P.~Protzel, ``Expectation-maximization for adaptive mixture models in graph optimization,'' in \emph{2019 international conference on robotics and automation (ICRA)}.\hskip 1em plus 0.5em minus 0.4em\relax IEEE, 2019, pp. 3151--3157.

\bibitem{neri2021approximate}
J.~Neri, P.~Depalle, and R.~Badeau, ``Approximate inference and learning of state space models with laplace noise,'' \emph{IEEE Transactions on Signal Processing}, vol.~69, pp. 3176--3189, 2021.

\bibitem{bakker1987tyre}
E.~Bakker, L.~Nyborg, and H.~B. Pacejka, ``Tyre modelling for use in vehicle dynamics studies,'' \emph{SAE Transactions}, pp. 190--204, 1987.

\bibitem{sutton1999policy}
R.~S. Sutton, D.~McAllester, S.~Singh, and Y.~Mansour, ``Policy gradient methods for reinforcement learning with function approximation,'' \emph{Advances in neural information processing systems}, vol.~12, 1999.

\end{thebibliography}

\newpage

 




\vfill

\end{document}